\documentclass[reprint,showpacs,showkeys,amsmath,amssymb,floats,showkeys,preprintnumbers,superscriptaddress,aps,prb]{revtex4-1}

\pdfoutput=1

\usepackage{bm}
\usepackage{graphicx}
\usepackage{float}
\usepackage{color}
\begin{document}

\preprint{Wu et al; Elastic properties of cubic Heusler compounds.}

\title{A critical study of the elastic properties and stability of Heusler compounds: \newline
       Cubic Co$_{2}YZ$ compounds with $L2_{1}$ structure}

\author{Shu-Chun Wu}
\affiliation{Max Planck Institute for Chemical Physics of Solids, D-01187 Dresden, Germany}

\author{S.~Shahab Naghavi}
\affiliation{Department of Materials Science and Engineering, Northwestern University, Evanston, Illinois 60208, USA} 

\author{Gerhard~H. Fecher}
\email{fecher@cpfs.mpg.de}
\affiliation{Max Planck Institute for Chemical Physics of Solids, D-01187 Dresden, Germany}

\author{Claudia Felser} 
\affiliation{Max Planck Institute for Chemical Physics of Solids, D-01187 Dresden, Germany}

\date{\today}

%%%%%%%%%%%%%%%%%%%%%%%%%%%%%%%%%%%%%%%%%%%%%%%%%%%%%%%%%%%%%%%%%%%%%%%%%%%%%%%%
\begin{abstract}

Elastic constants and their derived properties of various cubic Heusler
compounds were calculated using first-principles density functional
theory. To begin with, Cu$_2$MnAl is used as a case study to explain the
interpretation of the basic quantities and compare them with experiments. The
main part of the work focuses on Co$_2$-based compounds that are Co$_2$Mn$M$
with the main group elements $M=$~Al, Ga, In, Si, Ge, Sn, Pb, Sb, Bi, and
Co$_2TM$ with the main group elements Si or Ge, and the $3d$ transition metals
$T=$~Sc, Ti, V, Cr, Mn, and Fe. It is found that many properties of Heusler
compounds correlate to the mass or nuclear charge $Z$ of the main group element.

Blackman's and Every's diagrams are used to compare the elastic properties of
the materials, whereas Pugh's and Poisson's ratios are used to analyze the
relationship between interatomic bonding and physical properties. It is found
that the {\it Pugh's criterion} on brittleness needs to be revised whereas 
{\it Christensen's criterion} describes the ductile--brittle transition of Heusler
compounds very well. The calculated elastic properties give hint on a metallic
bonding with an intermediate brittleness for the studied Heusler compounds.

The universal anisotropy of the stable compounds has values in the range of
$0.57 <A_U <2.73$. The compounds with higher $A_U$ values are found close to the middle
of the transition metal series. In particular, Co$_2$ScAl with $A_U=0.01$ is
predicted to be an isotropic material that comes closest to an ideal Cauchy solid
as compared to the remaining Co$_2$-based compounds. Apart from 
the elastic constants and moduli, the sound velocities, Debye temperatures, and
hardness are predicted and discussed for the studied systems. The calculated
slowness surfaces for sound waves reflect the degree of anisotropy of the
compounds.

\end{abstract}
%%%%%%%%%%%%%%%%%%%%%%%%%%%%%%%%%%%%%%%%%%%%%%%%%%%%%%%%%%%%%%%%%%%%%%%%%%%%%%%%

\maketitle

%%%%%%%%%%%%%%%%%%%%%%%%%%%%%%%%%%%%%%%%%%%%%%%%%%%%%%%%%%%%%%%%%%%%%%%%%%%%%%%%
\section{Introduction}
 
There is a broad interest in Heusler compounds owing to the multitude of
different thermal, electrical, magnetic, and transport properties that are
realized in a rather simple crystalline structure. Owing to both applications
and fundamental interests, such as superconductivity, heavy fermions, the Kondo
effect, the Hall effect, and half-metallic ferromagnetism, these compounds are
among the most studied materials~\cite{gfp11}. Regular Heusler compounds
crystallize in a cubic {\it fcc} lattice with the space group $F\:m\bar{3}m$. In
certain cases, the cubic phases of regular Heusler compounds undergo, a
martensite-austenite phase transition to a tetragonal lattice, whereby the
symmetry changes to $I4/mmm$. In fact, both the cubic and the tetragonal phases
have attracted considerable attention owing to their half-metallic ferromagnetic
and spin-transfer-torque applications~\cite{CPR06,WBF08,SWK12}. Knowledge of the
stability of each of these phases is crucial for industrial applications as well
as fundamental research.

New Heusler compounds have been suggested to be stable in many theoretical
works. However, in many cases it is experimentally found that it is not possible
to synthesize these compounds. A possible reason for this is that  not all
stability criteria are respected in theoretical calculations. In fact, mostly
all the used stability criteria are {\it necessary} but not {\it sufficient}.
This implies that a suggested compound that fulfills a particular criterion may
not exist as it possibly violates other criteria. One of these necessary, but
not sufficient, criteria is the total energy, or the energy of formation,
satisfying the condition $E_{\rm compound}<\sum E_{\rm Elements}$ for a compound
to be stable. Further, the formation energy of the suggested compound needs to
be the minimum on the {\it ``convex hull''} taking into account all the
competing phases. Otherwise, it would decompose into other compounds with lower
energies. For example, the appearance of different binaries (XY, XZ, or other
similar combinations) may lead to a lower total energy as compared to a single
ternary (X$_2$YZ), and thus, hinder the formation of a Heusler compound.

Another important criterion is the mechanical stability of a predicted
structure. According to Born~\cite{Bor39}, a necessary condition for the
thermodynamic stability of a crystal lattice is that the crystals have to be
mechanically stable against arbitrary (small) homogeneous deformations. In fact,
this is the main concept of elastic constants. Elastic constants provide
important information concerning the strength of materials, and often act as
stability criteria or order parameters in the study of the problem of structural
transformations~\cite{AEU99,KAC97,WYP93}. Further physical properties, such as
hardness, velocity of sound, Debye temperature, and melting point are also
related to the elastic constants~\cite{Gil01,Gil09,Pea97,Fine84}. The
information is not only essential requirements for industrial applications
but also for fundamental research. Examples of the latter case are the
superconducting and heavy fermion systems, in which a drastic change of elastic
constants and related properties have been obtained upon phase
transition~\cite{BWF94,LKG89}.

There are several reports on calculations of elastic constants and phase
stability~\cite{OUKC08,SGS12,KCY11}. Members of the series of cubic Co$_2TM$
($T=$~transition element, $M=$~main group element) Heusler compounds have been
studied previously to some extent by various
authors~\cite{CPR06,CUC13,OFB11,YNC13}. In many cases, only the three
independent elastic constants and the bulk modulus are calculated. In only a few
cases, experiments were carried out to measure the hardness and melting
temperatures, and to compare them with the calculations. However, most works
have been carried out for specific cases, and almost not all relevant properties
have been calculated and compared with experiments or properties of other
compounds.

The present report is intended to investigate the mechanical properties of a
variety of Co$_2$-based Heusler compounds by calculating their elastic
constants. To begin with,  Cu$_2$MnAl, as a typical Heusler compound, is studied
to explain the basic quantities and their interpretation. Results for
Co$_2$Mn$M$ ($M=$ main group element), and Co$_2TM$ ($T=3d$-metal, $M=$~Al, Si)
are listed and discussed with an in-depth analysis of the physical properties
and chemical bonding. As elastic constants are derived from the second
derivative of the energy with respect to the lattice displacements, the use of
an accurate energy calculator is crucial. Here, the full-potential all-electron
method was used to calculate the elastic constants, and related properties. The
relationship between the interatomic bonding and the physical properties is
considered using Pugh's and Poisson's ratios. The Blackman's diagram provides
complementary information about the bonding character of the Heusler compounds.
A covalent to metallic bonding with an intermediate ductility or brittleness is
found for the studied Heusler compounds. Several other physical properties have
been extracted from the elastic constant calculations.

The present work concentrates on the half-metallic Co$_2$-based Heusler
compounds that have a high impact on magnetoelectronics. The results for the
elastic properties of tetragonal and phase change materials that exhibit
magnetic shape memory and magnetocaloric effects will be published
elsewhere~\cite{WNF17}. Some basic calculational aspects, including the
convergence of the method, are also found in Reference~\cite{WNF17}.

%%%%%%%%%%%%%%%%%%%%%%%%%%%%%%%%%%%%%%%%%%%%%%%%%%%%%%%%%%%%%%%%%%%%%%%%%%%%%%%%
\section{Computational details}

The elastic constants are second derivatives of the total energy with respect to
various lattice deformations. Therefore, accurate calculation of the total
energy is required. The full-potential linearized augmented plane wave (FLAPW)
technique is one such method that provides the required level of numerical
accuracy, albeit at the cost of complexity. In particular, in the case of
Heusler compounds, FLAPW is a reliable choice as some Heusler compounds are
sensitive to the employed method~\cite{FFe13}, and many Heusler compounds
contain atoms from the lanthanide or actinide series with occupied $f$-orbitals.
In the present work, the electronic structure was calculated using the full-potential 
linear augmented plane wave method, as implemented in {\sc Wien}2k~\cite{BSM01,SBl02}. 
The details of the calculations are reported in
References~\cite{KFF07b,FCF13} and a forthcoming publication~\cite{WNF17}. 
The charge density and other site specific properties were analyzed using
Bader's quantum theory of atoms in molecules (QTAIM)~\cite{Bad90} using the
built-in routines of {\sc Wien}2k as well as the {\sc Critic}2 package of
programs~\cite{OBP09,OJL14}. We developed our own routines, and used them to
determine the elastic constants and to analyze them in detail.

The Perdew--Burke--Ernzerhof implementation of the generalized gradient
approximation (PBE-GGA) was used for the exchange-correlation potential. The
number of plane waves was defined by $R_{\rm MT}k_{\max} = 9.0$ and 8000 $k$
points within the first Brillouin zone were used for integration. The energy
convergence criterion was set to 10$^{-5}$ Ry and the charge convergence was
less than a 10$^{-3}$ electronic charge in every case. The convergence of the
elastic constants with the parameters of the calculation has been already
reported in the publication on tetragonal compounds~\cite{WNF16}.

Most Co$_2$-based Heusler compounds are among the half-metallic ferromagnetic
materials, thus only ferromagnetic ordering has been studied here. See
References~\cite{KFF07b,FFe13} for details of the electronic and magnetic
structure of the investigated compounds. The basic equations for the
calculations of the elastic constants are  discussed in the following. More
details are provided in Appendix. The bulk moduli and relaxed lattice parameters
are found by fitting the calculated energy--volume relation to the Birch--Murnaghan 
equation of state~\cite{Bir47,Mur44}.

There are numerous ways to apply different strains and their combinations to the
crystal. For cubic crystals, there are three independent elastic constants and only
two more calculations are needed besides the bulk modulus that is found from the
equation of state. A necessary supplementary condition in the calculation of the elastic
constants is the conservation of volume when strain (or stress) is applied.
Table~\ref{tab:strain} summarizes the applied strains used in the present work
to determine the elastic constants. The applied strains are illustrated in
Figure~\ref{fig:straintypes}. More details are found in Reference~\cite{WNF16}
and in the Appendix.

% Table 2 %%%%%%%%%%%%%%%%%%%%%%%%%%%%%%%%%%%%%%%%%%%%%%%%%%%%%%%%%%%%%%%%%%%%%%
\begin{table*}[htb]
\centering
\caption{Applied strains to cubic structure are listed here. \\
         The isotropic strain type (0) is not volume conserving, 
         it is not directly used 
         but used together with the lattice parameter optimization 
         that yields the bulk modulus. The consequences of the strain 
         types are sketched in Figure~\ref{fig:straintypes}.}
    \begin{ruledtabular}
    \begin{tabular}{ll llll l}
    Type & Strain &  &  &  &  & $\Delta E / V_{0}$ \\
    \hline
    (0) & isotropic    & $e_{1}=\delta$                    & $e_{2}=\delta$  & $e_{3}=\delta$                                  & & $(c_{11}+2c_{12})\delta^{2}$, see bulk modulus! \\
    (1) & tetragonal   & $e_{1}=\delta$                    & $e_{2}=\delta$  & $e_{3}=-\delta \: (2+\delta)/(1+\delta)^{2}$    & & $(c_{11}-c_{12})\delta^{2}+O(\delta^{3})$\\
    (2) & orthorhombic & $e_{1}=\delta$                    & $e_{2}=-\delta$ & $e_{3}=\delta^{2}/(1-\delta^{2})$               & & $(c_{11}-c_{12})\delta^{2}+O(\delta^{4})$\\
    (3) & monoclinic   & $e_{1}=\delta^{2}/(1-\delta^{2})$ & &                                                 & $e_{4}=2\delta$ & $c_{44}\delta^{2}+O(\delta^{4})$ 
\\
\end{tabular}
    \end{ruledtabular}
\label{tab:strain}
\end{table*}
%%%%%%%%%%%%%%%%%%%%%%%%%%%%%%%%%%%%%%%%%%%%%%%%%%%%%%%%%%%%%%%%%%%%%%%%%%%%%%%%

%%%%%%%%%%%%%%%%%%%%%%%%%%%%%%%%%%%%%%%%%%%%%%%%%%%%%%%%%%%%%%%%%%%%%%%%%%%%%%%%
\begin{figure}[htb]
\centering
\includegraphics[width=8.5cm]{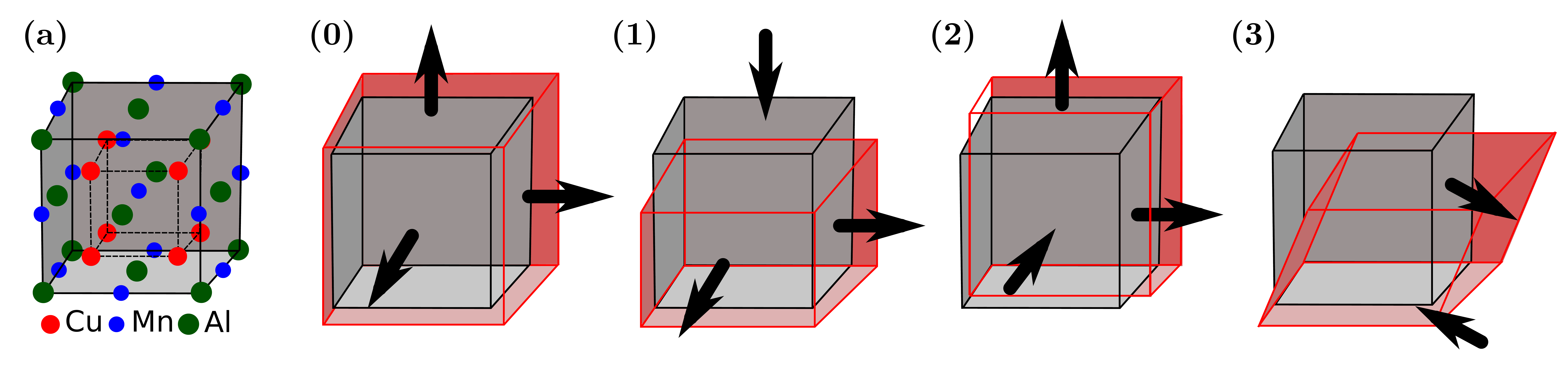}
 \caption{(Online color) Strain types for calculation of the elastic constants in cubic systems. \newline 
          (a) shows the cubic Heusler structure with $F\:m\overline{3}m$ symmetry.
          (0)...(3) sketch the strain types and resulting distortions according to Table~\ref{tab:strain}.
          Please note that (1) results in $x=y\neq z$ whereas in (2) all three lattice parameters 
          are different: $x\neq y\neq z$. See Table~\ref{tab:strain} for details of the strain types.}
\label{fig:straintypes}
\end{figure}
%%%%%%%%%%%%%%%%%%%%%%%%%%%%%%%%%%%%%%%%%%%%%%%%%%%%%%%%%%%%%%%%%%%%%%%%%%%%%%%%

In the present calculations, four distortions of each type in the range of $-3\%
\leq x \leq +3\%$ were applied to the relaxed structure with $V_0$ from the
structural optimization, which is the equation of state fitted to $E_{tot}(V)$. The
different distortions are sketched in Figure~\ref{fig:straintypes}. The energies
$E(x)$ of the monoclinic, orthorhombic and tetragonal strain types were fitted
to a $4^{th}$ order polynomial $E(x)=E_0 + a_2x^2 + a_3 x^3 + a_4 x^4$. Finally,
the $B$ and the $a_2$ values were used to determine the elastic constants and their
derived quantities. The elastic constants reported below are averaged over the
values determined by applying tetragonal or orthorhombic strains. The equations for
the properties calculated from the elastic constants are given in detail in 
Appendix~\ref{appendix}. It should be noted that models used for the Vicker's hardness 
($H{_V\rm^C}$) and the melting temperatures ($T_m^c$) are only suitable for cubic
structures~\cite{CNL11,Fine84} and may be used for the comparison of different
compounds rather than yielding absolute values.

%%%%%%%%%%%%%%%%%%%%%%%%%%%%%%%%%%%%%%%%%%%%%%%%%%%%%%%%%%%%%%%%%%%%%%%%%%%%%%%%
\section{Results}

To begin with, the results for the classical Heusler compound Cu$_2$MnAl
are presented and compared with experiments, because it is one of the few Heusler
compounds for which measured values of the elastic constants are available.
Table~\ref{tab:cu2mnal_elast} compares the calculated elastic properties with the
experimental work of Michelutti {\it et al}~\cite{MBL78}. As seen in
Table~\ref{tab:cu2mnal_elast}, the calculated elastic constants agree well with
the experiment, where the overestimation of about 10\% that is observed could  be due to the intrinsic
properties of the calculational method~\cite{LSO14} or due to uncertainties of
the experimental set-up. The other calculated properties, such as elastic moduli,
Cauchy pressure, and velocity of sound show excellent agreement with those found
in the experiment.

%%%%%%%%%%%%%%%%%%%%%%%%%%%%%%%%%%%%%%%%%%%%%%%%%%%%%%%%%%%%%%%%%%%%%%%%%%%%%%%%
\begin{table}[htb]
\centering
\caption{ Elastic properties of Cu$_2$MnAl.\\
          The experimental and optimized lattice parameters ($a$) are given in {\AA}.
          Elastic moduli $B$ (bulk), $G$ (shear), $E$ (Young), hardness parameter $H$, elastic constants $c_{ij}$, $C'$,
          and Cauchy pressure $p_{\rm C}$ are given in GPa. $k$, 
          $\nu$, $\zeta$, $A_e$ and $A_U$ are dimensionless quantities. Experimental values for the $c_{ij}$ are 
          taken from Reference~\cite{MBL78}.
          Values for 0~K are extrapolated from the temperature dependence shown in 
          Figure~3 of Reference~\cite{MBL78}. }
  \begin{ruledtabular}
  \begin{tabular}{ll|c|cc}
                    &      & Calculated & \multicolumn{2}{c}{Experiment} \\
                    &      &            & 0~K     & 300~K  \\
      \hline               
      $a$           & \AA  & 5.934      &         & 5.9615 \\                                            
%      $B_{BM}$ \\                      
%      $B'_{BM}$ \\                     
      \hline                                                                                                             
       $c_{11}$     & GPa  & 143.7      & 128.1   & $(136.1\pm0.3)$ \\
       $c_{12}$     & GPa  & 116.1      & 101.5   &  $(96.8\pm0.8)$ \\
       $c_{44}$     & GPa  & 117.6      & 104.4   &  $(94.1\pm0.2)$ \\
      \hline                            
       $C'$         & GPa  &  27.6      &  26.6   &  39.3    \\
       $p_{\rm C}$  & GPa  &  -1.5      &  -1.3   &  2.7     \\
      \hline                                                
       $B$          & GPa  & 125.3      & 110.4   &  109.9   \\
       $G$          & GPa  &  52.5      &  47.9   &   50.9   \\
       $E$          & GPa  & 138.0      & 125.6   &  132.2   \\
      \hline                                               
       $H$          & GPa  &   6.43     &   6.06  &    6.8   \\
      \hline                                               
       $k$          &      &   2.41     &  2.30   &   2.16   \\
       $\nu$        &      &   0.32     &  0.31   &   0.30   \\ 
       $\zeta$      &      &   0.866    &         &          \\
       $A_e$        &      &   9.64     &  7.85   &   4.79   \\
       $A_U$        &      &   7.97     &  7.17   &   3.6    \\
  \end{tabular}                
  \end{ruledtabular}
\label{tab:cu2mnal_elast}
\end{table}
%%%%%%%%%%%%%%%%%%%%%%%%%%%%%%%%%%%%%%%%%%%%%%%%%%%%%%%%%%%%%%%%%%%%%%%%%%%%%%%%

\begin{figure}[htb]
\centering
\includegraphics[width=8.5cm]{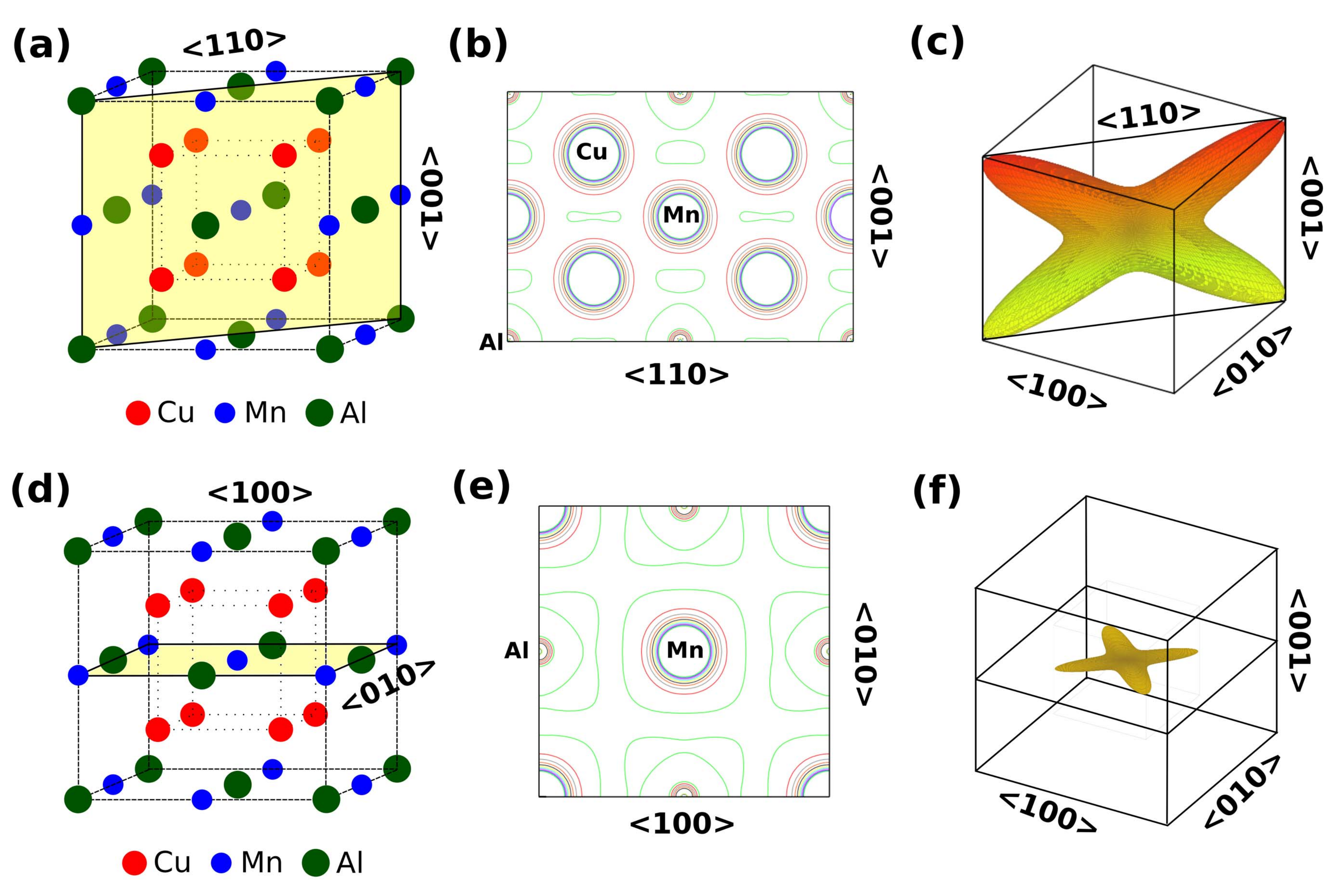}
   \caption{(Online color) Crystalline, electronic, and elastic structure of Cu$_2$MnAl. \\
            Compared are
            (a) Cubic structure with $(110)$ plane highlighted. 
            (b) The electron charge density plot in the $(110)$ plane. 
            (c) The $(110)$ cross-sections of Young's modulus. 
            (d) Cubic structure with $(001)$ plane highlighted. 
            (e) The electron charge density plot in the $(001)$ plane. 
            (f) The $(001)$ cross-sections of Young's modulus. }
\label{fig:cu2mnal}
\end{figure}

The elastic constants of Cu$_2$MnAl are listed in Table~\ref{tab:cu2mnal_elast}.
$c_{11}$, which represents stiffness against principal strains, is higher than
$c_{44}$, which represents shear deformation.  The shear ($G$) and tetragonal
shear $c'$ moduli are also low compared to the bulk modulus. This implies that 
Cu$_2$MnAl has the lowest resistance against shear deformations. The 
cross-sections on (110) and (001) crystallographic planes of Young's moduli of
Cu$_2$MnAl are shown in Figure~\ref{fig:cu2mnal}(c) and~(f). It is seen that the
anisotropy of Young's modulus is noticeable in both the planes. The directions where
the maxima appear correspond to the high-fracture energy directions, which are
along $\left<111\right>$ in the (110) plane and $\left<110\right>$-direction in
the (001) plane.

% Figure 2 %%%%%%%%%%%%%%%%%%%%%%%%%%%%%%%%%%%%%%%%%%%%%%%%%%%%%%%%%%%%%%%%%%%%%
%%%%%%%%%%%%%%%%%%%%%%%%%%%%%%%%%%%%%%%%%%%%%%%%%%%%%%%%%%%%%%%%%%%%%%%%%%%%%%%%

The Kleinman's parameter $\zeta$ describes the relative positions of the atoms under
strain. The calculated value of $\zeta\approx0.9$ for Cu$_2$MnAl suggests that the
atomic positions are rather rigid against distortions of the lattice. The
tetragonal shear modulus $c'=C'/2\approx(13\ldots20)$~GPa is the smallest
modulus and thus it is the main constraint on stability. Both anisotropies,
the Zener ratio $A_e$ and universal anisotropy $A_U$, are about 8. This is a rather
large value and may suggest elastic instability of Cu$_2$MnAl in the $L2_1$
structure, as will be discussed in the following.

\begin{figure}[htb]
\centering
\includegraphics[width=8.5cm]{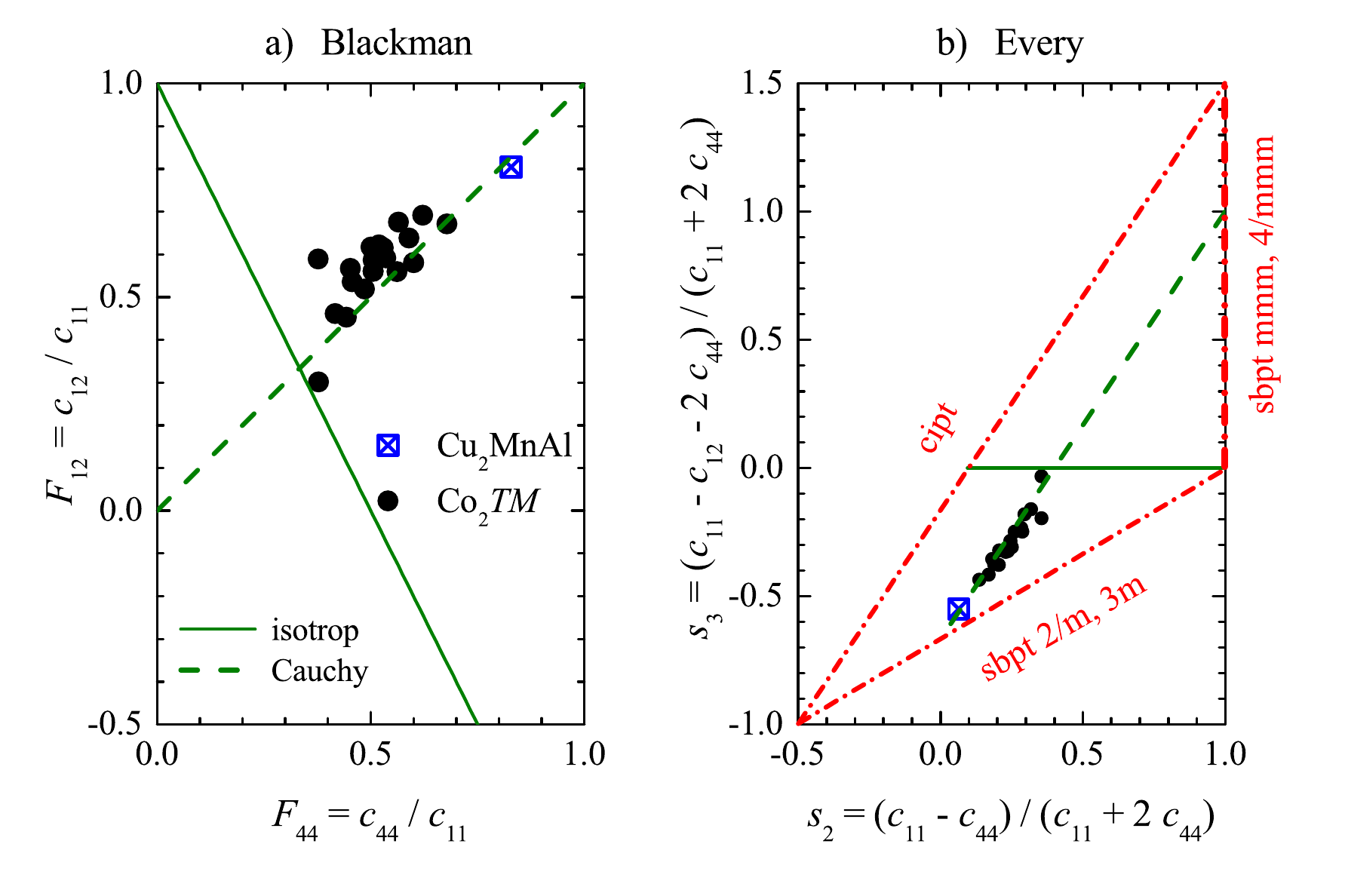}
   \caption{(Online color) (a) Blackman and (b) Every diagrams. \\ 
            The full line refers to the isotropic case, where $A_e=1$,
            the dashed line refers to the case where the Cauchy
            pressure $p\rm{_C}$ vanishes: $c_{44}=c_{12}$. 
            Ideal Cauchy solids appear at the point of intersection of the two lines.
            In Every's representation (b), the stability triangle according to the Born criteria 
            is marked by dash-dotted lines.
            Its three sides are given by the line of cubic isomorphous phase transitions (cipt)
            and two lines of symmetry breaking phase transitions (sbpt).}
\label{fig:black-every}
\end{figure}

The use of Blackman's and Every's diagrams~\cite{LMi08} is an efficient way to
compare the elastic properties of cubic materials. In both types of diagrams,
dimensionless quantities that are ratios of different moduli are correlated.
Figure~\ref{fig:black-every} summarizes the results of the present work in such
diagrams. Blackman's diagram compares in a simple way the ratios
$F_{ij}=c_{ij}/c_{11}$ of the elastic constants in a plot of $F_{12}(F_{44})$,
whereas Every's diagram is more complicated. It compares $s_3=(1-F_{12}-2
F_{44})/(1+2F_{44})$ as a function of $s_2=(1-F_{44})/(1+2 F_{44})$. Born's
shear criterion restricts $F_{44}$ to positive values and the spinodal and Born
criteria restrict $F_{12}$ to the range $-0.5<F_{12}<1$. Therefore, no
additional restrictions appear for Blackman's diagram and all values within the
entire range of Figure ~\ref{fig:black-every}(a) are allowed. In Every's
diagram, the values for stable systems need to fall into the triangle with
$(s_2,s_3)$ equal to (-1/2,-1)-(1,0)-(1,3/2) as is marked in Figure
~\ref{fig:black-every}(b).

% Figure 3 %%%%%%%%%%%%%%%%%%%%%%%%%%%%%%%%%%%%%%%%%%%%%%%%%%%%%%%%%%%%%%%%%%%%%
%%%%%%%%%%%%%%%%%%%%%%%%%%%%%%%%%%%%%%%%%%%%%%%%%%%%%%%%%%%%%%%%%%%%%%%%%%%%%%%%

From both the diagrams shown in Figure~\ref{fig:black-every}, it is seen that the
Heusler compounds calculated in the present work are close to the Cauchy line,
 where the Cauchy pressure vanishes. All the studied
compounds are in the region where the anisotropy index is positive. Obviously,
Cu$_2$MnAl has one of the highest anisotropies and comes close to the line of
$2/m,3m$ symmetry breaking phase transitions as marked in Every's diagram. The
Co$_2$-based compounds will be discussed below in more detail.

Table~\ref{tab:cu2mnal_prop} summarizes the physical properties of Cu$_2$MnAl that
are related or derived from the elastic properties. The measured~\cite{MBL78}
sound velocities at room temperature are in the range from 4553~m/s to 6003~m/s
for longitudinal modes depending on the direction of propagation and the
polarization. The average sound velocity in the transverse modes is about
$(3791\pm3)$~m/s. The values calculated with Anderson's approach~\cite{And63}
are in the same order of magnitude.

The Debye temperature was calculated from the mode averaged sound velocity
$\overline{v}$ and in the quasi-harmonic approach. Both methods result in values
of about 395~K. This is marginally higher  than the experimental value. The
slightly larger values are typical for the acoustical approaches that neglect
the optical phonons~\cite{OFB11,And63}.

%%%%%%%%%%%%%%%%%%%%%%%%%%%%%%%%%%%%%%%%%%%%%%%%%%%%%%%%%%%%%%%%%%%%%%%%%%%%%%%%
\begin{table}[htb]
\centering
\caption{ Properties of Cu$_2$MnAl. \\
          Sound velocities are calculated from elastic constants as given
          in Table~\ref{tab:cu2mnal_elast}. }
  \begin{ruledtabular}
  \begin{tabular}{ll|c|cc}
                            &           & Calculated & \multicolumn{2}{c}{Experiment} \\
                            &           &            & 0~K     & 300~K  \\
      \hline               
       $a$                  & \AA       & 5.934      &         & 5.9615  \\
       $\rho$               & kg/m$^3$  & 6637       &         & 6550    \\
       $m$                  &  g/mol    & \multicolumn{3}{c}{209.01}     \\
      \hline                                                                            
       $v_l$                & m/s       & 5421       &         &         \\
       $v_t$                & m/s       & 2807       & \multicolumn{2}{c}{see text} \\
       $\overline{v}$       & m/s       & 3142       &         &         \\ 
      \hline                                                                            
       $\Theta^{\rm ac}_D$  & K         &  397       &  372    &  376    \\
       $\zeta^{\rm ac}$     &           &    1.88    &         &         \\
       $f_\nu$              &           &    0.72    &  0.74   &  0.76   \\
       $\Theta^{\rm qha}_D$ & K         &  395       &         &         \\
      \hline                                                                            
       $T_m^c$              & K         & 1402       &         &         \\
       $H{_V\rm ^C}$        & GPa       &    4.38    &         &         \\
  \end{tabular}
  \end{ruledtabular}
\label{tab:cu2mnal_prop}
\end{table}
%%%%%%%%%%%%%%%%%%%%%%%%%%%%%%%%%%%%%%%%%%%%%%%%%%%%%%%%%%%%%%%%%%%%%%%%%%%%%%%%

\begin{figure*}[htb]
\centering
\includegraphics[width=15cm]{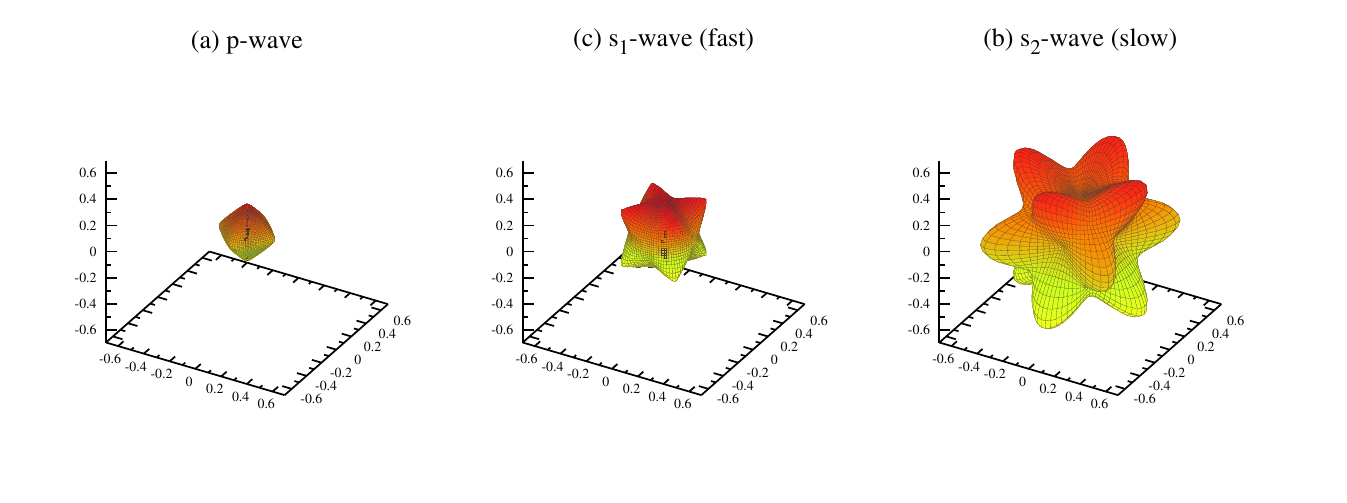}
   \caption{(Online color) Calculated slowness surfaces of Cu$_2$MnAl. \\
            The slowness is given in (km/s)$^{-1}$. }
\label{fig:slowness_cu2mnal}
\end{figure*}

Figure~\ref{fig:slowness_cu2mnal} shows the three sheets of the slowness surface
of Cu$_2$MnAl. Pronounced extrema, arising from the large anisotropy of the
compound, are observed for all the three modes. The pressure ($p$) wave has the
highest phase velocity. The minima of its slowness appear along \{111\}-type
directions that is along the space diagonals and the maxima are found along the
\{001\}-type, principle axes. The maxima of the slowness of the fast shear wave
($s_2$) are along the \{111\}-type directions. The slowness surface of the slow
shear wave ($s_1$) appears to be more complicated; its maxima appear along
\{110\}-type directions. The two shear modes are sixfold degenerate, that is,
their slowness is the same at the six $\left[ 00s \right ]$-type points along the three
principal axes.
 
% Figure 4 %%%%%%%%%%%%%%%%%%%%%%%%%%%%%%%%%%%%%%%%%%%%%%%%%%%%%%%%%%%%%%%%%%%%%
%%%%%%%%%%%%%%%%%%%%%%%%%%%%%%%%%%%%%%%%%%%%%%%%%%%%%%%%%%%%%%%%%%%%%%%%%%%%%%%%

One may roughly categorize materials as ductile (malleable) or as brittle with
respect to mechanical characteristics, as for example, machinability. For
various applications, the materials need to be malleable, for instance if they
are required to be used as wires. Ductile materials usually exhibit metallic
bonding, whereas high brittleness indicates a more covalent or ionic character
of the bonds. The transition region between these subjective criteria is
blurred. Because of the importance in various applications, the malleability
criteria are of great significance. Pugh's and Poisson's ratios are very helpful
mechanical parameters in the characterization of maleability (brittle or ductile).

\begin{figure}[htb]
\centering
\includegraphics[width=6cm]{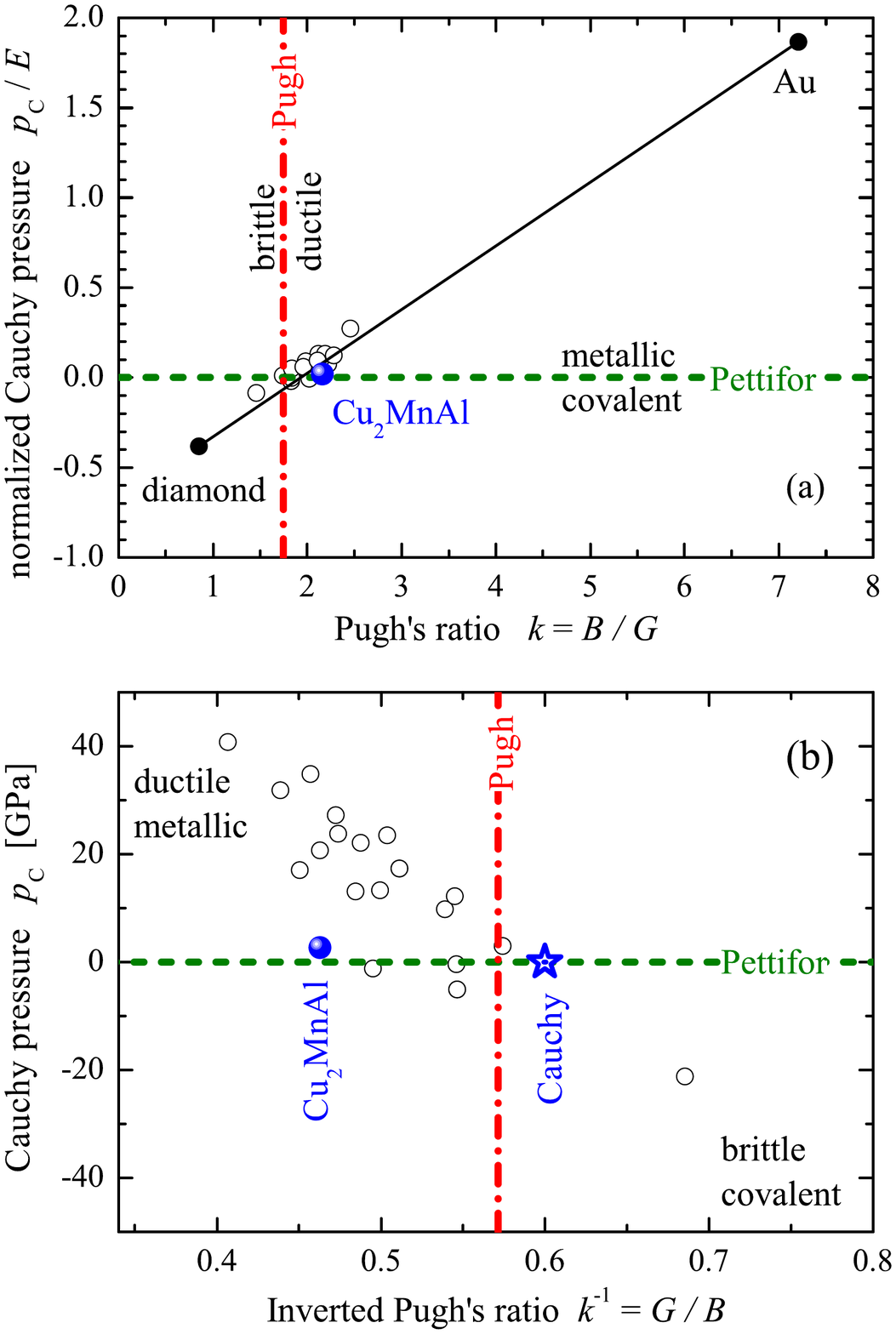}
   \caption{(Online color) Ductile-brittle diagram for various Co$_2$-based Heusler compounds. \\ 
            Values are given as open circles. The horizontal dashed and vertical 
            dash-dotted lines correspond to the metallic--covalent and 
            ductile-brittleness criteria of Petifor and Pugh, respectively.
            Cu$_2$MnAl is marked by a spherical symbol and an ideal Cauchy solid
            by an asterix. 
            The values for the extremal elements from experiments, fcc--Au as 
            {\it "most soft"} and diamond--C as {\it "most hard"}, 
            are given for comparison. The solid  line connecting them is drawn
            for better comparison. }
\label{fig:brittle}
\end{figure}

Pettifor~\cite{Pet92} proposed the criterion that a positive Cauchy pressure
indicates metallic bonds whereas negative Cauchy pressures are typical in the
case of covalent bonds. Another older criterion is based on Pugh's
work~\cite{Pug54}. According to the so-called {\it "Pugh's criterion"}, many
publications~\cite{KVZ11,VSD11,NCL12} indicate that the critical value 
($k_{\rm cr}$) that separates brittle ($k\leq k_{\rm cr}$) and ductile 
($k\geq k_{\rm cr}$) materials is around $1.75$ or $k^{-1}\approx0.571$. It is
worthwhile to mention that Pugh's ratio for a cubic, isotropic Cauchy solid is
$k_{\rm Cauchy}=5/3=1.6\overline{6}$, as shown in Figure~\ref{fig:brittle}(a).
The behavior of the Heusler compounds investigated in the present work is
summarized in Figure~\ref{fig:brittle} where the Cauchy pressure is plotted as a
function of Pugh's ratio. The two criteria are drawn as vertical and horizontal
lines. Following the value for Pettifor's and Pugh's criteria, most of the
studied compounds should be classified as ductile or metallic materials. There
is an obvious contradiction between the empirical rules and the observation that
the Co$_2$-based Heusler compounds are all brittle instead of  ductile. In
particular, it appears that Pugh's criterion needs to be modified.

% Figure 5 %%%%%%%%%%%%%%%%%%%%%%%%%%%%%%%%%%%%%%%%%%%%%%%%%%%%%%%%%%%%%%%%%%%%%
%%%%%%%%%%%%%%%%%%%%%%%%%%%%%%%%%%%%%%%%%%%%%%%%%%%%%%%%%%%%%%%%%%%%%%%%%%%%%%%%

Poisson's ratio $\nu$ is related to Pugh's ratio by~\cite{Nye85}:

\begin{equation}
   \nu = \frac{3k - 2}{6k + 2}.
\end{equation}

The valid range of Pugh's ratio ($0 < k \leq \infty$) restricts Poisson's ratio
to $-1 < \nu \leq 1/2$. The Poisson’ ratio $\nu_0=0$ is obtained  for
$k_0=2/3$. Materials with $\nu<0$ are called {\it ``auxcetic''}. Typically,
Poisson's ratio of covalent materials is small $\nu\approx0.1$, whereas it is
greater than  $0.33$ for metallic materials. Poisson's ratio indicates the
degree of directionality of the covalent bonds. Smaller Poisson's ratios
indicate a stronger degree of covalent bonding resulting in higher hardness. The
so-called {\it "Frantsevich rule"} is  widely used as a criterion for
brittleness, which is based on the tables of elastic properties in the book by
Frantsevich {\it et al}~\cite{FVB83}. According to this rule, compounds with a
Poisson ratio of $\nu_{cr}\leq0.33$ are brittle and those with
$\nu_{cr}\geq0.33$ are ductile or malleable. It should be noted that this value
is not given explicitly in Reference~\cite{FVB83}. It is based on properties
reported for the materials tabulated by Frantsevich and was later accepted as
empirical rule.

Referring to the original work~\cite{Pug54}, Pugh also did not suggest a
criterion. However, Pugh  only mentioned that Ir was the least malleable metal
($k=1.74$) and Au was the most malleable metal ($k=6.14$). Based on present
knowledge, it is obvious that Ir is hard and brittle~\cite{Ir}, and hence, the
critical value could possibly be between 1.74 and 6.14. Based on the relation
between $k$ and $\nu$, it follows that the critical Poisson's ratio
$\nu_{cr}\approx1/3$ of Frantsevich's rule corresponds to a critical Pugh's
ratio of $k_{\rm cr}\approx2.66$ ($k^{-1}_{\rm cr}\approx3/8$), so that the two
empirical rules only differ in the exact number that distinguishes between the
two types of behaviors.

Generally, it may be considered  that materials with $\nu=0$ are absolutely brittle,
whereas those with $\nu=1/2$ are perfectly ductile. Christensen~\cite{Chr13} used
the failure theory to describe the mechanical properties. He introduced a
nanoscale variable $\kappa$, which characterizes the relative size of the bond
bending and the bond stretching effects. Further, he related it to
renormalized Poisson's or Pugh's ratios and defined the ductility $D$ by:

\begin{eqnarray}
	D      & = & (1-\kappa)^2          \\ \nonumber
	\kappa & = & \frac{1-2\nu}{1+\nu} = \frac{2}{3k}.
\end{eqnarray}

According to this relation, materials with $D=0$ are absolutely brittle, those
with $D=1$ are perfectly ductile, and the brittle--ductile transition takes
place at $D_{B/D}=1/2$. The latter implies that

\begin{equation}
	\kappa_{B/D}=1-\sqrt{1/2},
\end{equation}

and thus, the critical values of the Poisson's and Pugh's ratios defined as,

\begin{eqnarray}
	\nu_{B/D} & = & \frac{1}{3\sqrt{2}-1}          \\ \nonumber
	k_{B/D}   & = & \frac{2}{3}\frac{\sqrt{2}}{\sqrt{2}-1}.
\end{eqnarray}

$\nu_{B/D}\approx0.31$ is close to $\nu_{cr}$ of Frantsevich's rule. Further, it
corresponds to a critical Pugh's ratio of $k_{B/D}\approx2.3$ 
($k_{B/D}^{- 1}\approx0.44$), that is clearly larger than the so-called Pugh's
criterion. It is worthwhile to note that the ductility of a Cauchy solid becomes
$D_{\rm Cauchy}=(3/5)^2=0.36$, which implies that Cauchy solids should be more
brittle than ductile.

\begin{figure}[htb]
\centering
\includegraphics[width=8.5cm]{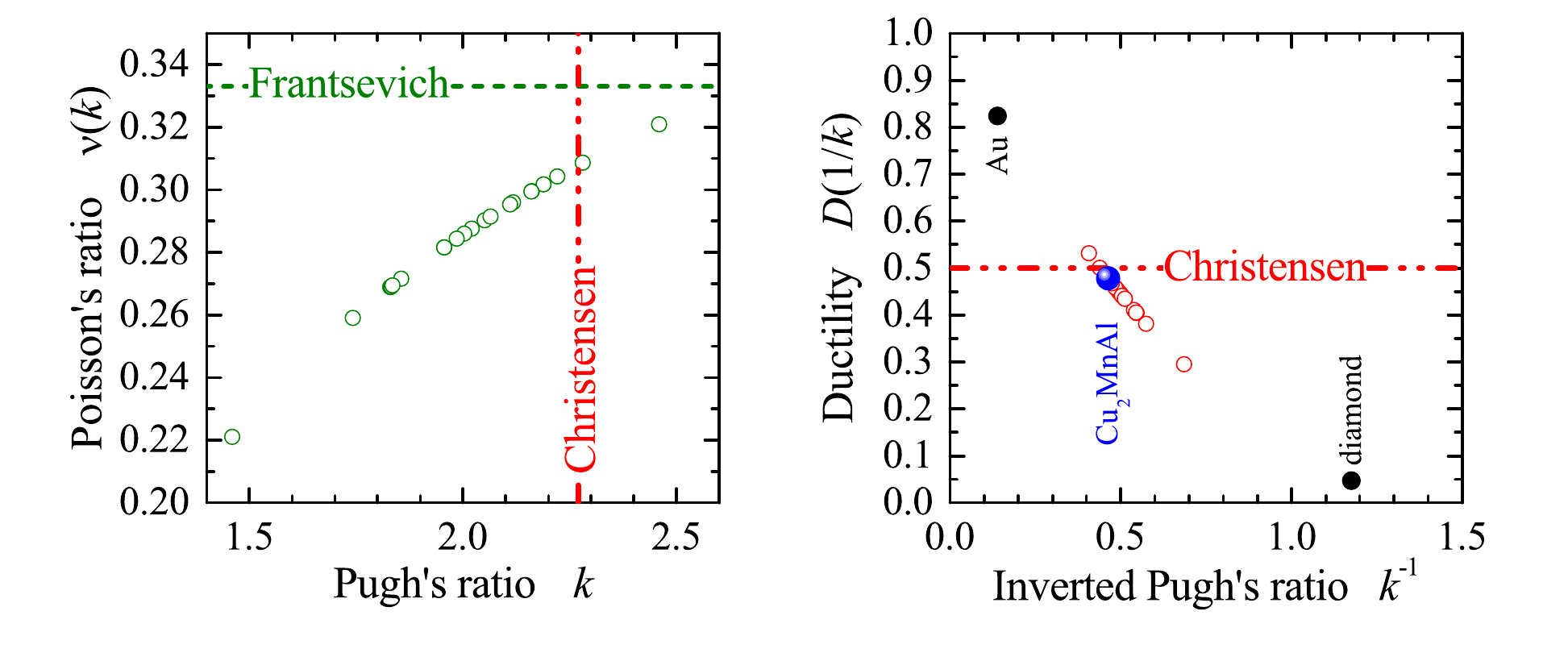}
   \caption{(Online color) Poisson's ratio and ductility for various Co$_2$-based Heusler compounds. \\ 
            Values are given as open circles. The horizontal dashed and 
            Dash-dotted lines correspond to the ductile-brittleness criteria of 
            Frantsevich and Christensen, respectively.
            Cu$_2$MnAl is denoted by a spherical symbol. Note that Poisson's ratio is plotted in the range 
            of the calculated compounds, whereas the ductility is shown in the entire intervals
            of $D$ and $k^{-1}$. }
\label{fig:ductile}
\end{figure}

The behavior of the Heusler compounds is shown in Figure~\ref{fig:ductile} that
relates Poisson's ratio and Christensen's ductility to Pugh's ratio. It is
obvious that the compounds ---and in particular Cu$_2$MnAl--- are far away from
the extreme elements, which are elemental Au, and C in the form of diamond, as the most
ductile and most brittle elements, respectively.

% Figure 6 %%%%%%%%%%%%%%%%%%%%%%%%%%%%%%%%%%%%%%%%%%%%%%%%%%%%%%%%%%%%%%%%%%%%%
%%%%%%%%%%%%%%%%%%%%%%%%%%%%%%%%%%%%%%%%%%%%%%%%%%%%%%%%%%%%%%%%%%%%%%%%%%%%%%%%

The above discussion and the experimental observations on various Heusler
compounds lead to the conclusion that Frantsevich's rule (or better
Christensen's ductility criterion) is suitable for the compounds studied here,
whose behaviors lie on the border-line between brittleness and ductility.

The crystalline structure of Cu$_2$MnAl is shown in Figure~\ref{fig:cu2mnal}(a)
and~(d). Figure~\ref{fig:cu2mnal}(b) and~(c) show the charge densities and
Young's modulus of Cu$_2$MnAl in the (110) plane. Figure~\ref{fig:cu2mnal}(e)
and~(f) show the charge densities and Young's modulus of Cu$_2$MnAl in the (001)
plane. Bader's QTAIM analysis was used to analyze the charge density and
magnetic moments. The results of the QTAIM analysis are listed in
Table~\ref{tab:qtaim}, where a charge transfer is observed. On the average, about 0.9
electrons are transferred from the Mn and Al atoms to the Cu atoms with relatively larger
contribution from the Al atoms. The Mn atoms carry a magnetic moment of
3.45~$\mu_B$, whereas Cu and Al exhibit only a negligible polarization.

%%%%%%%%%%%%%%%%%%%%%%%%%%%%%%%%%%%%%%%%%%%%%%%%%%%%%%%%%%%%%%%%%%%%%%%%%%%%%%%%
\begin{table}[htb]
\centering
\caption{  QTAIM analysis of Cu$_2$MnAl. \newline
            $V$ are the basin volumes $V$,
            $n_e$ is the number of electrons in the basin, and $q=Z-n_e$ is the 
            electron excess/deficiency with respect 
            to the electron occupation in free atoms $Z$.  
            Negative values indicate electron excess, that is, negatively charged ions, and
            $m$ is the magnetic moment in the basin in multiples 
            of the Bohr magneton. }
   \begin{ruledtabular}
   \begin{tabular}{l cccc}
                     & $V$ [{\AA}$^3$] & $n_e$   &  $q$   & $m$ [$\mu_B$]   \\
   \hline
      Cu             & 15.547          & 29.86   & -0.86  & 0.03  \\
      Mn             & 12.228          & 24.57   &  0.43  & 3.45  \\
      Al             &  8.969          & 11.71   &  1.28  & 0.04  \\
   \end{tabular}
   \end{ruledtabular}
\label{tab:qtaim} 
\end{table}
%%%%%%%%%%%%%%%%%%%%%%%%%%%%%%%%%%%%%%%%%%%%%%%%%%%%%%%%%%%%%%%%%%%%%%%%%%%%%%%%

The QTAIM critical points of Cu$_2$MnAl and their properties are summarized in
Table~\ref{tab:crit}. There are, indeed, three different nuclei that act as
attractors. The cage critical point $c$ is found between Mn and Al along the
[001] axis and acts as a repeller, which is the absolute minimum of the charge
density. Further, two bond critical points $b_{1,2}$ are  located
between Cu and Al ($b_1$), and between Cu and Mn ($b_2$). The third bond critical
point is located in between the Mn atoms along the [001] direction. When the two
ring critical points $r_{1,2}$ are also considered, the Morse sum of the numbers
$n_i$ of the different critical points vanishes ($n_n-n_b+n_r-n_s=0$), as
expected for crystals.

%%%%%%%%%%%%%%%%%%%%%%%%%%%%%%%%%%%%%%%%%%%%%%%%%%%%%%%%%%%%%%%%%%%%%%%%%%%%%%%%
\begin{table}[htb]
\centering
\caption{  QTAIM critical point analysis of Cu$_2$MnAl. \newline
            $pg$ is the point group symmetry of the critical point, and mult is the multiplicity
            of the critical points in the conventional cubic cell; the multiplicities
            in the primitive cell are mult/4. }
  \begin{ruledtabular}
  \begin{tabular}{ll ccc cr}
      $pg$     & type       &  \multicolumn{3}{c}{position}          &  mult  &  name     \\   %  f            |grad|        lap
  \hline
      $O_h$    & nucleus    &  0          & 0          & 0           &   4    & Mn        \\   %  1.33354E+04   0.00000E+00  -3.00000E+15
      $O_h$    & nucleus    &  1/2        & 1/2        & 1/2         &   4    & Al        \\   %  1.49069E+03   0.00000E+00  -3.00000E+15
      $T_d$    & nucleus    &  1/4        & 1/4        & 1/4         &   8    & Cu        \\   %  2.26397E+04   0.00000E+00  -3.00000E+15
      
      $C_{3v}$ & bond       &  0.397      & 0.397      & 0.397       &  32    & $b_1$     \\   %  3.59964E-02   2.30158E-17   8.03886E-03
      $C_{3v}$ & bond       &  0.878      & 0.878      & 0.878       &  32    & $b_2$     \\   %  4.07746E-02   3.31141E-17   2.32844E-02
      $D_{2h}$ & bond       &  1/4        & 1/4        & 1/2         &  24    & $b_3$     \\   %  2.93205E-02   4.92026E-17   3.70579E-02
      
      $C_{2v}$ & ring       &  0          & 0.302      & 0.302       &  48    & $r_1$     \\   %  2.88242E-02   1.12244E-17   3.50207E-02
      $C_{2v}$ & ring       &  0          & 0.287      & 0.213       &  48    & $r_2$     \\   %  2.92312E-02   1.36397E-16   3.39558E-02
      $C_{4v}$ & cage       &  0          & 0          & 0.254       &  24    & $c$       \\   %  2.79046E-02   2.80557E-13   2.07398E-02
  \end{tabular}
  \end{ruledtabular}
\label{tab:crit} 
\end{table}
%%%%%%%%%%%%%%%%%%%%%%%%%%%%%%%%%%%%%%%%%%%%%%%%%%%%%%%%%%%%%%%%%%%%%%%%%%%%%%%%

The analysis of the bonding type with the properties of the critical points is
discussed in Reference~\cite{MPL02}. Metallic systems exhibit a flat electron
density $\rho$ throughout the valence region. The flatness
$f=\rho^{c}_{\min}/\rho^{b}_{\max}$ is a measure of the metallicity of the compounds.
$\rho^{c}_{\min}$ is the cage critical point, at which, the density is minimum,
and $\rho^{b}_{\max}$ is the highest density among all the bond critical points.
For Cu$_2$MnAl, it is $f = 0.684$. This is of the same order of magnitude as the
flatness in Cu or Fe (both $\approx0.57$; see Reference~\cite{MPL02}), whereas
compounds with covalent bonding typically have $f$ ratios of less than 0.1. From
the large electronic flatness, the bonding in Cu$_2$MnAl is clearly metallic.

In the following, the results of the calculations for various Co$_2$-based
Heusler compounds are discussed. These compounds, containing Mn in particular, 
are of much interest in spintronic applications. The Mn containing
compounds are discussed in the first part. The second part discusses the
variation of the $3d$ transition metal in Co$_2TM$ when the main group element
$M$ is attached to Al or Si, and the $3d$ transition metal $T$ ($T=$~Sc, Ti, V, Cr,
Mn, Fe) is varied.

%%%%%%%%%%%%%%%%%%%%%%%%%%%%%%%%%%%%%%%%%%%%%%%%%%%%%%%%%%%%%%%%%%%%%%%%%%%%%%%%
\subsection{Results for Co$_2$Mn$M$ ($M=$ main group element)}

In this section, the elastic and mechanical properties of the Mn containing
Heusler compounds Co$_2$Mn$M$ ($M=$~Al, Ga, In, Si, Ge, Sn, Pb, Sb, Bi) are
discussed. Table~\ref{tab:epmn} compares the mechanical properties of the
Co$_2$Mn$M$ compounds. It may be noted that the compounds with  main group
heavy elements (In, Pb, Sb, Bi) have not been synthesized up to now and Heusler
compounds with this composition most probably do not exist. They are used here
to complete the trends when changing the main group elements. It should also be
noted that those compounds are stable --- at least from the Born-Huang criteria
(These criteria are discussed in the Appendix).

%%%%%%%%%%%%%%%%%%%%%%%%%%%%%%%%%%%%%%%%%%%%%%%%%%%%%%%%%%%%%%%%%%%%%%%%%%%%%%%%
\begin{table*}[htb]
\centering
\caption{ Elastic properties of Co$_2$Mn$M$ ($M=$ main group element).\\
 The experimental ($a_{\rm exp}$) and optimized ($a_{\rm opt}$) lattice parameters are given in {\AA}.
 Elastic moduli $B$, $G$, $E$, hardness parameter $H$, elastic constants $c_{ij}$, $C'$,
 and Cauchy pressure $p_{\rm C}$ are given in GPa. $k$, 
 $\nu$, $\zeta$, $A_e$, and $A_U$ are dimensionless quantities. }
 \begin{ruledtabular}
 \begin{tabular}{l|ccc|cccc|cc}
 $M$ & Al & Ga & In & Si & Ge & Sn & Pb & Sb & Bi \\
 \hline
 $a_{\rm exp}$ &5.755$^a$&5.770$^b$& &5.654$^b$&5.743$^b$&6.000$^b$& & & \\ 
 $a_{\rm opt}$ & 5.700 & 5.718 & 5.974 & 5.643 & 5.730 & 5.987 & 6.102 & 6.019 & 6.184 \\
% $B_{BM}$ \\
% $B'_{BM}$ \\
 \hline 
 $c_{11}$ & 267.3 & 242.2 & 194.4 & 310.5 & 272.8 & 233.6 & 200.1 & 235.3 & 192.0 \\
 $c_{12}$ & 155.3 & 167.5 & 130.6 & 174.2 & 160.0 & 138.3 & 124.5 & 133.6 & 113.2 \\
 $c_{44}$ & 160.4 & 150.5 & 131.8 & 156.9 & 137.9 & 125.0 & 103.8 & 106.4 & 72.4 \\
 \hline
 $C'$ & 112.0 & 74.7 & 63.8 & 136.3 & 112.8 & 95.3 & 75.6 & 101.7 & 78.8 \\
 $p_{\rm C}$ & -5.1 & 17.0 & -1.2 & 17.3 & 22.1 & 13.3 & 20.7 & 27.2 & 40.8 \\
 \hline 
 $B$ & 192.6 & 192.4 & 151.9 & 219.6 & 197.6 & 170.0 & 149.7 & 167.5 & 139.5 \\
 $G$ & 105.3 & 86.6 & 75.2 & 112.3 & 96.4 & 85.0 & 69.3 & 79.1 & 56.7 \\
 $E$ & 267.2 & 225.9 & 193.6 & 287.9 & 248.7 & 218.5 & 180.1 & 205.1 & 149.8 \\
 \hline
 $H$ & 16.2 & 11.3 & 10.7 & 16.4 & 13.5 & 12.1 & 8.0 & 10.8 & 6.8 \\
 \hline 
 $k$ & 1.83 & 2.22 & 2.02 & 1.96 & 2.05 & 2.00 & 2.16 & 2.12 & 2.46 \\
 $\nu$ & 0.27 & 0.30 & 0.29 & 0.28 & 0.29 & 0.29 & 0.30 & 0.30 & 0.32 \\ 
 $\zeta$ & 0.692 & 0.779 & 0.764 & 0.676 & 0.696 & 0.701 & 0.767 & 0.681 & 0.699 \\
 $A_e$ & 2.86 & 4.05 & 4.14 & 2.30 & 2.44 & 2.62 & 2.75 & 2.09 & 1.84 \\
 $A_U$ & 1.46 & 2.73 & 2.85 & 0.88 & 1.02 & 1.21 & 1.88 & 0.68 & 0.46 \\
 \end{tabular}
 \end{ruledtabular}
\leftline{$^a$Ref.\cite{UKF08}
$^b$Ref.\cite{BNW00}}
\label{tab:epmn}
\end{table*}
%%%%%%%%%%%%%%%%%%%%%%%%%%%%%%%%%%%%%%%%%%%%%%%%%%%%%%%%%%%%%%%%%%%%%%%%%%%%%%%%

The relaxed lattice parameters $a_{\rm opt}$ agree well with the experimental
values.  They exhibit the typical trend that the lattice parameter increases
with the nuclear charge ($Z$) of the main group element. The elastic constants
of the Co$_2$Mn$M$ compounds follow the general inequality 
$B > c_{44} > G > c' > 0$. 
Here, the isotropic shear or rigidity modulus $G$ is not the main
constraint on stability. The smallest values are obtained for $c'=C'/2$, that is
the tetragonal shear modulus is the limiting parameter for the stability of the
cubic $L2_1$ structure of the investigated compounds. The values of $c'$ are in
the range of 37--68~GPa and thus far above the values required to force
tetragonal instabilities in the vicinity of ambient pressure. The lowest values
for the bulk moduli are obtained for the compounds containing  the main group,
heavy elements, which  exhibit large lattice parameters. The Pb and Bi
containing compounds exhibit comparably low values of the rigidity moduli.
Within each group, the values of the hardness parameter decrease with increasing
$Z$ of the main group elements. As compared to the previous
calculations~\cite{CUC13}, the elastic constants and bulk modulus fit quite well
for most compounds, and the differences are lesser than 20~GPa for each
quantity.

Pugh's ratio $k$ of the Co$_2$Mn$M$ compounds ranges from about 1.83 to 2.46
with a mean value of $\overline{k}\approx 2.1$. Similarly, Poisson's ratio $\nu$
falls also in a narrow range 0.27--0.32 with a mean value of
$\overline{\nu}\approx 0.29$. A convenient way of quantifying the degree of 
off-axis anisotropy in the elastic constants for a cubic crystal is to use the Zener
ratio. Here, Zener ratio ($A_{e}$) exhibits no trend of a dependency on the main
group element. It exhibits the smallest value for Co$_2$MnBi (1.84) and the
largest value for Co$_2$MnIn (4.14). The latter value is exceptional (compare
with the next section) and points to a large elastic anisotropy in the material.
Another method is the usage of the so-called universal anisotropy index
($A_{U}$), which shows the same tendency as the Zener ratio. Especially for
Co$_2$MnSb, in spite of the value of the anisotropy being in a reasonable range,
this compound does not exist, and tested samples exhibit phase
separation~\cite{KMW06}. The anisotropy does only judge on the structural
stability but not on the chemical stability.

% Figure 7 %%%%%%%%%%%%%%%%%%%%%%%%%%%%%%%%%%%%%%%%%%%%%%%%%%%%%%%%%%%%%%%%%%%%%
\begin{figure}[htb]
\centering
\includegraphics[width=8.5cm]{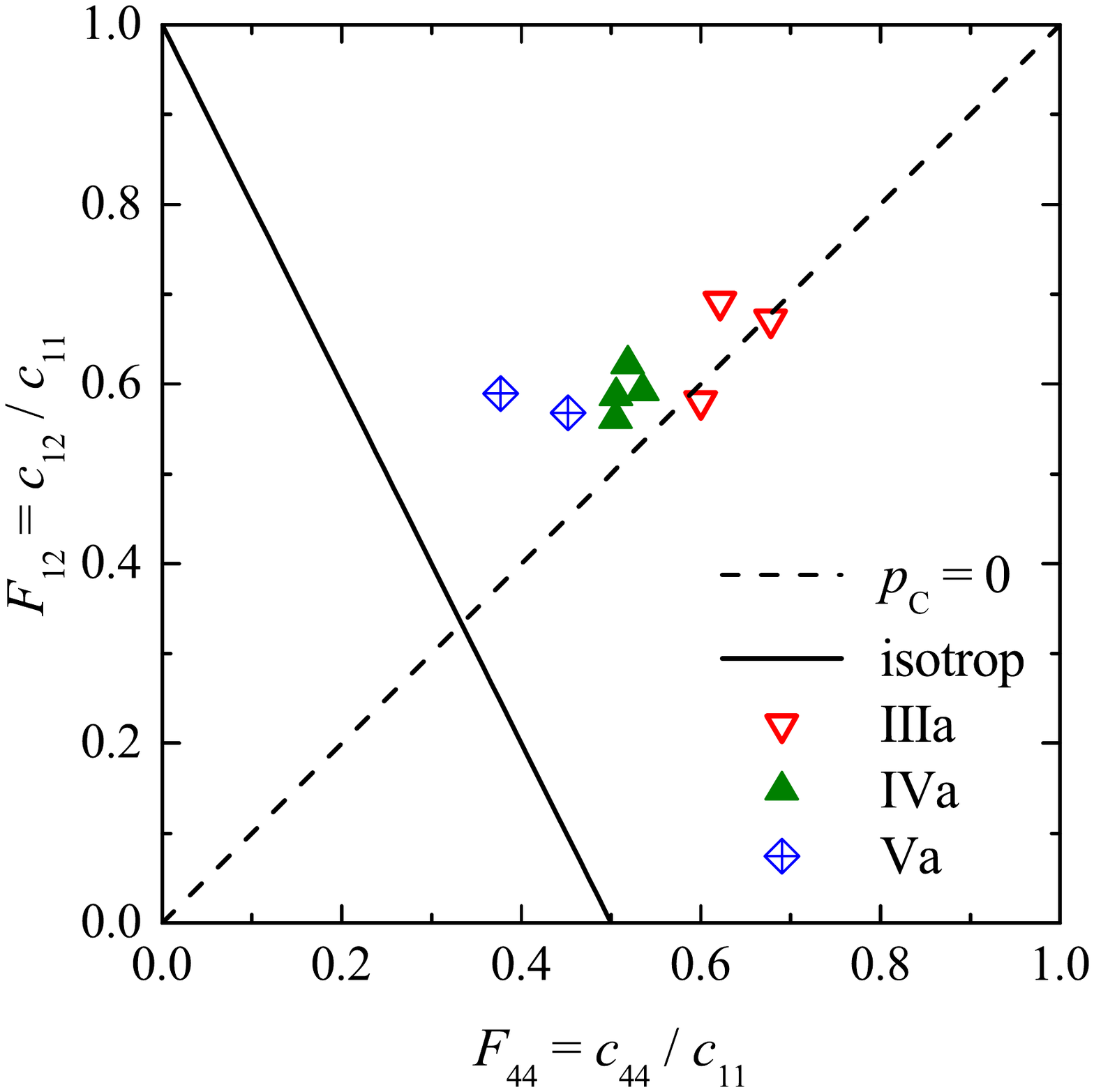}
 \caption{(Online color) Blackman diagram for Heusler compounds based on Co and Mn. \\
 The full line is for the isotropic case, where $A_e=1$,
 the dashed line is for  the limit where the Cauchy
 pressure $p\rm{_C}$ vanishes: $c_{44}=c_{12}$. Different symbols
 are used for the main group elements from different groups:
 IIIa (Al, Ga, In), IVa (Si, Ge, Sn, Pb), Va (Sb, Bi)}
\label{fig:bdmn}
\end{figure}
%%%%%%%%%%%%%%%%%%%%%%%%%%%%%%%%%%%%%%%%%%%%%%%%%%%%%%%%%%%%%%%%%%%%%%%%%%%%%%%%

Figure~\ref{fig:bdmn} shows the Blackman's diagram using the elastic data of the
Co$_2$Mn$M$ Heusler compounds given in Table~\ref{tab:epmn}. All $F$ ratios fall
in the allowed range for mechanical stability. The values of $F_{12}$ appear
close together in a region around the Cauchy line where the Cauchy pressure
vanishes. They also fall in the region of positive Zener ratios ($A_e>1$). The
figure suggests the type of bonding, covalent or metallic. A positive Cauchy
pressure is suggestive of greater degree of metallic bonding. On the contrary,
when the Cauchy pressure is negative, there appears to be greater degree of
covalent bonding. As discussed in the foregoing  and in agreement with the
Poisson's ratio, all the studied compounds are metallic and on the borderline
between brittleness and ductility. This is in agreement with experiments, in which
all these compounds have a silvery metallic luster.

\begin{figure}[htb]
\centering
\includegraphics[width=8.5cm]{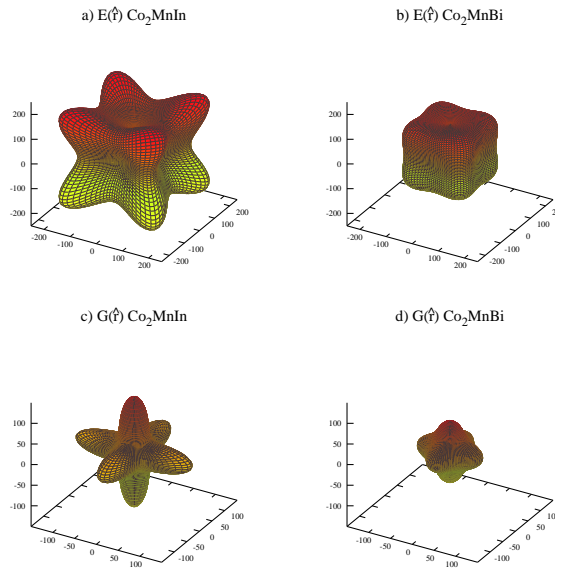}
   \caption{(Online color) Calculated distribution of Young's $E(\hat{r})$ 
            and rigidity $G(\hat{r})$ moduli
            of Co$_2$MnIn and Co$_2$MnBi. }
\label{fig:ygm1}
\end{figure}

In fact, directional dependent plots of rigidity $G(\hat{r})$ and Young's
$E(\hat{r})$ moduli are an alternative visual way of showing the Zener
anisotropy. The implication of the elastic anisotropy on the elastic moduli will
be illustrated for the two borderline cases with the largest (In) and the smallest (Bi)
anisotropy. The three dimensional distributions of $G(\hat{r})$ and $E(\hat{r})$
are shown in Figure~\ref{fig:ygm1} for Co$_2$MnIn and Co$_2$MnBi.

% Figure 8 %%%%%%%%%%%%%%%%%%%%%%%%%%%%%%%%%%%%%%%%%%%%%%%%%%%%%%%%%%%%%%%%%%%%%
%%%%%%%%%%%%%%%%%%%%%%%%%%%%%%%%%%%%%%%%%%%%%%%%%%%%%%%%%%%%%%%%%%%%%%%%%%%%%%%%

The anisotropy of Young's modulus of the In and Bi containing compounds is
displayed in Figures~\ref{fig:ygm1}(a) and~(b) that show the three dimensional
distribution $E(\hat{r})$. The pronounced anisotropy of the In containing
compound is clearly visible. Figures~\ref{fig:ygm1}(c) and~(d) show the three
dimensional distribution $G(\hat{r})$ of the rigidity moduli of the two
compounds. Again, the differences in the anisotropy of the moduli are clearly
visible. Comparing the distribution of the moduli, it is obvious that Young's
modulus is largest in the $\left<111\right>$-type directions whereas the
rigidity modulus is largest in the $\left<100\right>$-type directions, that is
along the cubic axes. This behavior is generic for  all compounds listed in
Table~\ref{tab:epmn} and a direct consequence of the condition $A_e>1$. These anisotropic
compounds exhibit different responses to stress or strain when tested in
different directions. The anisotropic property is particularly important for
applications where mechanical stress is applied to the materials, directly or by
thermal expansion and contraction.

%%%%%%%%%%%%%%%%%%%%%%%%%%%%%%%%%%%%%%%%%%%%%%%%%%%%%%%%%%%%%%%%%%%%%%%%%%%%%%%% 
\subsection{Results for Co$_2TM$ ($T=3d$-metal, $M=$ Al, Si)}

In this section, the influence of the $3d$ transition metal on the elastic and
mechanical properties of selected Co$_2$-based Heusler compounds is discussed.
Table~\ref{tab:epalsi} compares the elastic properties of the Co$_2TM$ ($T=$~Sc,
Ti, V, Cr, Mn, Fe and $M=$~Al, Si) compounds. Al and Si were selected as the main
group elements, because they exhibit the most complete series over the $3d$
transition metals that exist in reality. The Sc compounds as well as Co$_2$CrSi
have not been synthesized up to now and do not possibly exist. They are used here
to complete the trends of the properties when changing the transition metal. Similar to the
compounds with heavy main group elements reported above, those compounds are
stable --- at least from the Born-Huang criteria.

%%%%%%%%%%%%%%%%%%%%%%%%%%%%%%%%%%%%%%%%%%%%%%%%%%%%%%%%%%%%%%%%%%%%%%%%%%%%%%%%
\begin{table*}[htb]
\centering
\caption{ Elastic properties of Co$_2TM$ ($T=$ Sc to Fe and $M=$ Al, Si).\\
          The experimental ($a_{\rm exp}$) and optimized ($a_{\rm opt}$) lattice parameters are given in {\AA}.
          Elastic moduli $B$, $G$, $E$, hardness parameter $H$, elastic constants $c_{ij}$, $C'$,
          and Cauchy pressure $p_{\rm C}$ are given in GPa. $k$, 
          $\nu$, $\zeta$, and $A_e$ are dimensionless quantities. }
  \begin{ruledtabular}
  \begin{tabular}{l | cccccc | cccccc}
      $M$     & &      & Al & & & & &                                             & Si & & &  \\
      $T$              & Sc     &    Ti   &    V    &    Cr   &    Mn   &    Fe   & Sc     &    Ti   &    V    &  Cr    &  Mn     &   Fe   \\
      \hline 
      $a_{exp}$        &        &5.848$^c$&5.780$^d$&5.726$^e$&5.755$^a$&5.730$^f$&        &5.740$^c$&5.657$^g$&        &5.654$^b$&5.640$^h$\\
      $a_{opt}$        &  5.972 &  5.837  &  5.758  &  5.711  &  5.700  &  5.702  &  5.870 &  5.758  &  5.679  &  5.638 &  5.643  &  5.630 \\
      \hline                                                                                                         
       $c_{11}$        & 277.8  & 286.0   &  290.0  &  265.0  & 267.3   &  268.2  & 267.4  &  295.4  & 297.4   & 287.5  & 310.5  & 273.6   \\
       $c_{12}$        &  83.8  & 129.7   &  150.5  &  169.2  & 155.3   &  150.1  & 123.5  &  158.3  & 183.5   & 194.2  & 174.2  & 168.5   \\
       $c_{44}$        & 105.0  & 126.7   &  140.7  &  156.1  & 160.4   &  150.5  & 111.3  &  134.8  & 148.7   & 162.4  & 156.9  & 144.7   \\
      \hline
       $C'$            & 194.0  & 156.3   &  139.5  &   95.8  & 112.0   &  118.1  & 143.9  &  137.1  & 113.9   & 103.3  & 136.3  & 105.1\\
       $p_{\rm C}$     & -21.2  &   3.0   &    9.8  &   13.1  &  -5.1   &   -0.4  &  12.2  &   23.5  &  34.8   &  31.8  &  17.3  &  23.8\\       
      \hline                                                                                                                    
       $B$             & 148.4  & 181.8   &  197.0  &  201.5  &  192.6  &  189.5  & 171.5  &  204.0  & 221.5   & 225.3  & 219.6  & 203.5   \\
       $G$             & 101.7  & 104.4   &  106.2  &   97.2  &  105.3  &  103.5  &  93.4  &  102.8  & 101.2   &  98.8  & 112.3  &  96.4   \\
       $E$             & 248.4  & 262.9   &  270.0  &  251.3  &  267.2  &  262.6  & 237.2  &  264.0  & 263.5   & 258.6  & 287.9  & 249.8  \\
      \hline                                                                                                                    
       $H$             &  18.9  &  16.8   &   16.2  &   13.5  &   16.2  &   15.9  &  14.4  &   14.8  &  13.4   &  12.6  &  16.4  &  13.1   \\
      \hline                                                                                                                   
       $k$             &   1.46 &   1.74  &    1.86 &    2.07 &   1.83  &   1.83  &   1.84 &   1.99  &   2.19  &   2.28 &   1.96 &   2.11  \\
       $\nu$           &   0.22 &   0.26  &    0.27 &    0.29 &   0.27  &   0.27  &   0.27 &   0.28  &   0.30  &   0.31 &   0.28 &   0.30  \\
       $\zeta$         &   0.449&   0.585 &    0.641&    0.738&   0.692 &   0.675 &   0.593&   0.655 &   0.721 &   0.767&   0.676& 0.720\\
       $A_e$           &   1.08 &   1.62  &    2.02 &    3.28 &   2.86  &   2.55  &   1.55 &   1.97  &   2.61  &   3.48 &   2.30 &   2.76  \\
       $A_U$           &   0.01 &   0.29  &    0.62 &    1.88 &   1.46  &   1.13  &   0.23 &   0.57  &   1.19  &   1.75 &   0.88 & 1.34  \\
  \end{tabular}
  \end{ruledtabular}
\leftline{$^a$Ref.\cite{UKF08}
$^b$Ref.\cite{BNW00}
$^c$Ref.\cite{WZi73}
$^d$Ref.\cite{KCE10}
$^e$Ref.\cite{KUO08}
$^f$Ref.\cite{BEJ83}
$^g$Ref.\cite{FSS94}
$^h$Ref.\cite{WFK06}}
\label{tab:epalsi}
\end{table*}
%%%%%%%%%%%%%%%%%%%%%%%%%%%%%%%%%%%%%%%%%%%%%%%%%%%%%%%%%%%%%%%%%%%%%%%%%%%%%%%%

The elastic constants of the Co$_2TM$ compounds follow the general inequality 
$B > c_{44} > G > C'/2 > 0$ 
as was also observed above for the Mn-containing
compounds with varying main group elements. As in the earlier case, the
tetragonal shear modulus $c'=C'/2$ is the most critical of the moduli for
crystal stability. The bulk moduli are slightly greater in the Si containing
compounds as compared to the Al- containing compounds. The Young's and
rigidity moduli fall in the ranges (237.2--287.9)~GPa and (93.4--112.3)~GPa,
respectively. Our calculated values of the bulk moduli and elastic constants for
Co$_{2}T$Si (except $T$ = Sc) agree well with those reported by Chen {\it et
al}~\cite{CPR06}. Only the value of $c_{11}$ of Co$_2$VSi exhibits  a large
deviation of 20\%, whereas all others deviate by less than 7\%.

\begin{figure}[htb]
\centering
\includegraphics[width=8.5cm]{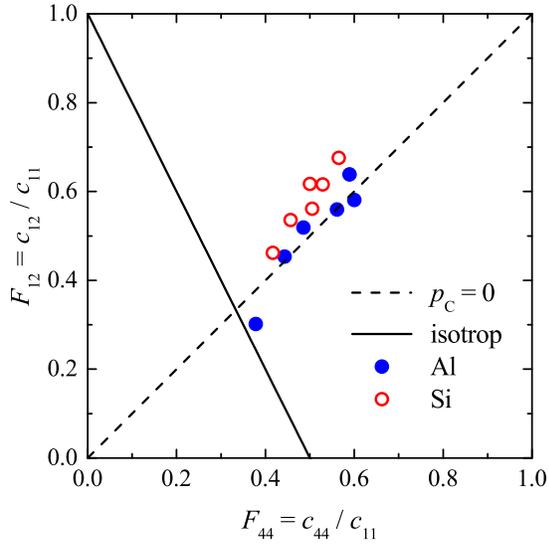}
   \caption{(Online color) Blackman diagram for Co$_2TM$ Heusler compounds containing 
            a $3d$ transition metal $T$ and Al or Si as the main group element $M$.}
\label{fig:bdalsi}
\end{figure}

As observed above for the Mn-containing compounds, all Co$_2TM$ compounds are
between brittle and ductile from Frantsevich's rule based on Poisson's ratio
($\nu_{cr}\leq0.33$) and Pugh's criterion ($k_{cr}\leq1.75$). The Co$_2$-based
compounds synthesized in our laboratories turned out to be mostly brittle in
accordance with the prediction of our calculations. Further discussion on the
elastic properties will be presented using the Blackman's diagram that is
shown in Figure~\ref{fig:bdalsi} for the Co$_2TM$ Heusler compounds. The values
of $F_{12}$ fall in a very narrow range about the Cauchy line. All Si-containing
compounds exhibit a positive Cauchy pressure, whereas $p\rm{_C}<0$ for the Al compounds 
with Sc, Mn, or Fe. It is worthwhile to note that the Al-containing
Co$_2$ compounds tend to antisite disorder, that is, they exhibit a $B2$ type
rather than a $L2_1$ type crystalline structure. It is interesting to note that
the hypothetical compound Co$_2$ScAl is assumed  to be nearly isotropic and very
similar to an ideal Cauchy solid. Its universal anisotropy of only 1\% is remarkable.
Even though the values of $c_{44}/c{11}$, $k$, and $\nu$ still deviate from the
ideal Cauchy values by about 10\%, out of all the compounds investigated here, it is 
closest to a Cauchy solid.

% Figure 9 %%%%%%%%%%%%%%%%%%%%%%%%%%%%%%%%%%%%%%%%%%%%%%%%%%%%%%%%%%%%%%%%%%%%%
%%%%%%%%%%%%%%%%%%%%%%%%%%%%%%%%%%%%%%%%%%%%%%%%%%%%%%%%%%%%%%%%%%%%%%%%%%%%%%%%

The elastic Zener (universal) anisotropy ranges from 1.08 (0.01) for Co$_2$ScAl
to 3.48 (1.88) for Co$_2$CrSi (Co$_2$CrAl). Comparing the elastic anisotropy of
compounds that are well known to crystallize in an ordered $L2_1$ structure and
those that are known to tend to disorder or where no successful synthesis is
reported up to now, the Zener ratios for the most stable compounds  are in the
range  $1.62<A_e<2.86$.

\begin{figure}[htb]
\centering
\includegraphics[width=8.5cm]{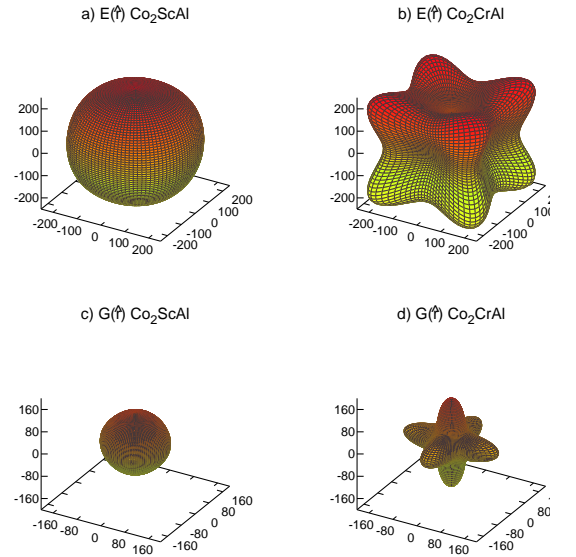}
   \caption{(Online color) Calculated distribution of Young's $E(\hat{r})$ 
            and rigidity $G(\hat{r})$ moduli
            of Co$_2$ScAl and Co$_2$CrAl. }
\label{fig:ygm2}
\end{figure}

\begin{figure}[htb]
\centering
\includegraphics[width=8.5cm]{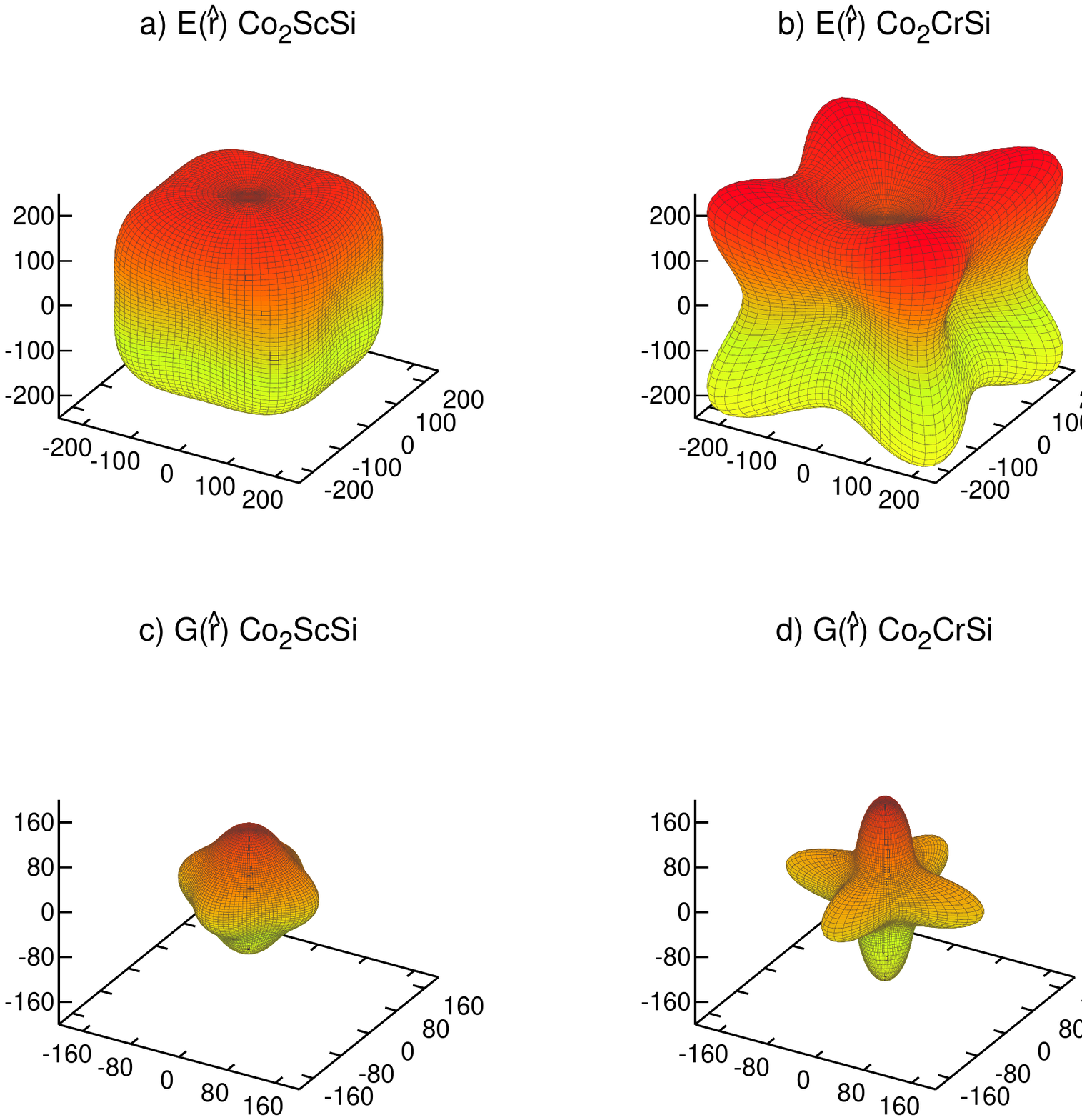}
   \caption{(Online color) Calculated distribution of Young's $E(\hat{r})$ 
            and rigidity $G(\hat{r})$ moduli
            of Co$_2$ScSi and Co$_2$CrSi. }
\label{fig:ygm3}
\end{figure}

The anisotropy of Young's moduli and the rigidity moduli of Co$_2$ScAl,
Co$_2$CrAl, Co$_2$ScSi and Co$_2$CrSi are displayed in
Figures~\ref{fig:ygm2} and~\ref{fig:ygm3} that show the three dimensional
distribution $E(\hat{r})$ and $G(\hat{r})$ as was also plotted above for the Mn
containing compounds. The more pronounced anisotropy of the Cr containing
compound is clearly visible. The  differences in the anisotropy of the
moduli between the two compounds are clearly visible, as observed above.

Comparing the distribution of the moduli, the Young's modulus is largest in the
$\left<111\right>$-type directions, whereas the rigidity modulus is largest in
the $\left<100\right>$-type directions for most compounds, which are listed in
Table~\ref{tab:epalsi}. The only exception is Co$_{2}$ScAl, where $A_e$ is close
to unity and thus the distributions are nearly spherical, which shows a tendency to
distortion in the $\left<100\right>$ direction.

% Figure 10 %%%%%%%%%%%%%%%%%%%%%%%%%%%%%%%%%%%%%%%%%%%%%%%%%%%%%%%%%%%%%%%%%%%%%
%%%%%%%%%%%%%%%%%%%%%%%%%%%%%%%%%%%%%%%%%%%%%%%%%%%%%%%%%%%%%%%%%%%%%%%%%%%%%%%%

% Figure 11 %%%%%%%%%%%%%%%%%%%%%%%%%%%%%%%%%%%%%%%%%%%%%%%%%%%%%%%%%%%%%%%%%%%%%
%%%%%%%%%%%%%%%%%%%%%%%%%%%%%%%%%%%%%%%%%%%%%%%%%%%%%%%%%%%%%%%%%%%%%%%%%%%%%%%%

%%%%%%%%%%%%%%%%%%%%%%%%%%%%%%%%%%%%%%%%%%%%%%%%%%%%%%%%%%%%%%%%%%%%%%%%%%%%%%%%
\section{Derived properties}
\label{prop}

This section summarizes the physical properties of the compounds that are
derived from the calculated elastic constants.  Table~\ref{tab:phy1} summarizes
the results for the Co$_2$-based, Mn containing Heusler compounds with variation
of the main group element Co$_2$Mn$M$ ($M=$ main group element). Properties of
the Co$_2$-based Heusler compounds Co$_2TM$ with varying $3d$ transition metals
$T$, and Al and Si as the main group elements $T$ are summarized in
Table~\ref{tab:phy2}

%%%%%%%%%%%%%%%%%%%%%%%%%%%%%%%%%%%%%%%%%%%%%%%%%%%%%%%%%%%%%%%%%%%%%%%%%%%%%%%%
\begin{table*}[htb]
\centering
\caption{ Physical properties of Co$_2$Mn$M$ ($M=$ main group element). 
The Debye temperature $\Theta^{\rm ac}_D$ is estimated from the sound velocity $\overline{v}$. }
  \begin{ruledtabular}
  \begin{tabular}{ll|ccc|cccc|cc}
      $M$                   &           & Al     & Ga      & In      & Si      & Ge      & Sn      & Pb     & Sb      & Bi      \\
      \hline 
       $\rho$               & kg/m$^3$  & 7166   & 8617    &   8959  &   7424  &   8665  &   9023  &  11110 &  8972   &  10726  \\
       $m$                  &  g/mol    & 199.8  & 242.5   &   287.6 &   200.9 &   245.4 &   291.5 &  380.0 &  294.6  &   381.8 \\
      \hline                                                                            
       $v_l$                & m/s       & 6817   & 5977    &   5305  &   7054  &   6135  &   5604  &  4668  &   5516  &   4478  \\
       $v_t$                & m/s       & 3833   & 3170    &   2897  &   3890  &   3335  &   3068  &  2497  &   2970  &   2299  \\
       $\overline{v}$       & m/s       & 4265   & 3543    &   3231  &   4335  &   3720  &   3421  &  2789  &   3316  &   2575  \\ 
      \hline                                                                            
       $\Theta^{\rm ac}_D$  & K         &  561   &  465    &    406  &   576   &    487  &   429   &   343  &   413   &    312  \\
       $\zeta^{\rm ac}$     &           & 1.59   & 1.80    &   1.70  &  1.66   &   1.71  &  1.69   &  1.77  &  1.75   &   1.91  \\
      \hline                                                                            
       $T_m^c$              & K         & 2133   & 1984    &   1702  &  2388   &   2165  &   1934  &  1736  &   1944  &   1688  \\
       $H{_V\rm ^C}$        & GPa       & 12.0   &  7.7    &    8.0  &  11.4   &    9.5  &    8.9  &   6.7  &    7.7  &    4.4  \\
  \end{tabular}
  \end{ruledtabular}
\label{tab:phy1}
\end{table*}
%%%%%%%%%%%%%%%%%%%%%%%%%%%%%%%%%%%%%%%%%%%%%%%%%%%%%%%%%%%%%%%%%%%%%%%%%%%%%%%%

The Debye temperature $\Theta^{\rm ac}_D$ and Gr{\"u}neisen parameters
$\zeta^{\rm ac}$ are estimated in the acoustical approximation from the mode
averaged sound velocities~$\overline{v}$. These quantities depend, in addition
to the elastic constants, on the mass density of the materials. The other two
properties, melting temperature and hardness are exclusively based on the
elastic constants. The melting temperature $T_m^c$ is roughly estimated from
$c_{11}$, and the hardness $H_V\rm ^C$ results from Pugh's and rigidity moduli.
The underlying ideas and equations are given in the Appendix. Table~\ref{tab:phy1} 
and Table~\ref{tab:phy2} summarize the properties derived for
various Co$_2TM$ Heusler compounds. In addition, the density $\rho$ and the
molecular mass $m$ are given for completeness. Interestingly, the remaining
physical properties do not appear to depend much on the composition. However,
some clear trends are recognized on  closer inspection.

Table~\ref{tab:phy1} reveals, for the Mn containing compounds, the trend that the
sound velocities calculated from the elastic constants and the Debye temperature
in the acoustical approach decrease with $Z$ of the main group element, whereas
the Gr{\"u}neisen parameters are all nearly the same for the different
compounds. The average values of $\zeta^{ac}$ is about 1.7. The acoustical Debye
temperatures are in the range of 312~K--576~K with average values of about
$(440\pm70)$~K. The range of validity for the melting temperature $T_m^c$ is
$\pm300$~K for the approximation used here, and thus is on the same order as that of the 
spread of the calculated values. This leads to the estimate that the melting
temperatures of the compounds are about ($1960\pm200$)~K. The Vickers's hardness
has average values of the order of ($8.5\pm1.8)$~GPa.

From Table~\ref{tab:phy2}, it is found that the sound velocities calculated from
the elastic constants exhibit only small changes among the different transition
metals. As a direct consequence, the values for the Debye temperature or
Gr{\"u}neisen parameters in the acoustical approach are all nearly the same for
the different compounds. There is no evidence of a distinguishable dependence on
the $T$ element,  as already mentioned in the previous sections. The
$\zeta^{ac}$ values of Co$_2$ScAl, Co$_2$TiAl, Co$_2$TiSi,  Co$_2$MnSi and
Co$_2$ScSi, however, are slightly below those of the remaining compounds that
exhibit average values of 1.57 for Al and 1.72 for Si. The acoustical Debye
temperatures are in the range of 533~K--576~K with average values of
$(565\pm10)$~K and $(555\pm10)$~K for the Al and Si compounds, respectively. The
similarity of the acoustical parameters arise from similar masses of Al and Si
that determines to a large extent, the vibrational properties of the compounds.
The range of validity for the melting temperature $T_m^c$ is $\pm300$~K for the
 approximation used here and is of the same  order as the spread of the
calculated values. This leads to the estimate that the melting temperatures
should be of the order of ($2180\pm50$)~K for the Al containing compounds
and ($2260\pm60$)~K for the Si containing compounds. All calculated melting temperatures are
consistently larger than  the experimental values~\cite{YNC13}. On the average, the Al
containing compounds exhibit larger hardness values as compared to the  Si
containing compounds. In both these groups, the Cr containing compounds exhibit
clearly lower values for the calculated hardness as compared to the other
transition metals. Neglecting the Cr values, the hardness exhibits average values
of ($12.4\pm1.4)$~GPa for the Al and $(9.9\pm1.1)$~GPa for the Si containing compounds.
The only two reported experimental values of hardness for Co$_2$MnGe and Co$_2$MnSi are
known to be 7.3 and 7.9, respectively~\cite{OFB11}. Although the experimental
values are smaller than the predicted values, the tendency is the same. This is expected
from the approximate nature of the model used.

%%%%%%%%%%%%%%%%%%%%%%%%%%%%%%%%%%%%%%%%%%%%%%%%%%%%%%%%%%%%%%%%%%%%%%%%%%%%%%%%
\begin{table*}[htb]
\centering
\caption{ Physical properties of Co$_{2}TM$ ($T=$~Sc, Ti, V, Cr, Mn, Fe; $M=$ Al, Si). 
The Debye temperature $\Theta^{\rm ac}_D$ is estimated from the sound velocity $\overline{v}$. }
  \begin{ruledtabular}
  \begin{tabular}{ll | cccccc | cccccc}
      $M$                   & & &       & Al & & & & &                                             & Si & & &  \\
      $T$                   &           & Sc     &    Ti   &    V    &    Cr   &    Mn   &    Fe   & Sc     &    Ti   &    V    &  Cr    &  Mn     &   Fe   \\
      \hline 
       $\rho$               & kg/m$^3$  & 5919   & 6436    &   6814  &   7019  &   7166  &   7189  &  6269  &  6742   &   7142  &  7336  &  7424   & 7511  \\
       $m$                  &  g/mol    & 189.8  & 192.7   &   195.8 &   196.8 &   199.8 &   200.7 &  190.9 &  193.8  &   196.9 &  197.9 &  200.9  & 201.8 \\
      \hline                                                                            
       $v_l$                & m/s       & 6928   & 7062    &   7049  &   6868  &   6817  &   6748  &  6872  &   7112  &   7064  &  6978  &  7054   & 6649  \\
       $v_t$                & m/s       & 4145   & 4027    &   3947  &   3722  &   3833  &   3794  &  3860  &   3904  &   3764  &  3670  &  3890   & 3583  \\
       $\overline{v}$       & m/s       & 4587   & 4476    &   4394  &   4153  &   4265  &   4221  &  4295  &   4352  &   4206  &  4104  &  4335   & 4000  \\ 
      \hline                                                                            
       $\Theta^{\rm ac}_D$  & K         &  576   &  575    &    572  &   546  &    561  &   555   &   549  &   567    &    556  &   546  &   576   &  533  \\
       $\zeta^{\rm ac}$     &           &  1.37  &  1.54   &    1.61 &   1.73 &    1.59 &   1.60  &   1.60 &   1.68   &    1.78 &   1.83 &   1.66  &  1.74 \\
      \hline                                                                            
       $T_m^c$              & K         & 2195   & 2244    &   2267  &  2119  &   2133  &   2138  &  2133  &   2299   &   2311  &   2252 &  2388   & 2170  \\
       $H{_V\rm ^C}$        & GPa       & 16.2   & 12.9    &   11.9  &   9.4  &   12.0  &   11.9  &  11.0  &   10.5   &    8.9  &    8.2 &  11.4   &   9.1 \\
  \end{tabular}
  \end{ruledtabular}
\label{tab:phy2}
\end{table*}

\begin{figure*}[htb]
\centering
\includegraphics[width=15cm]{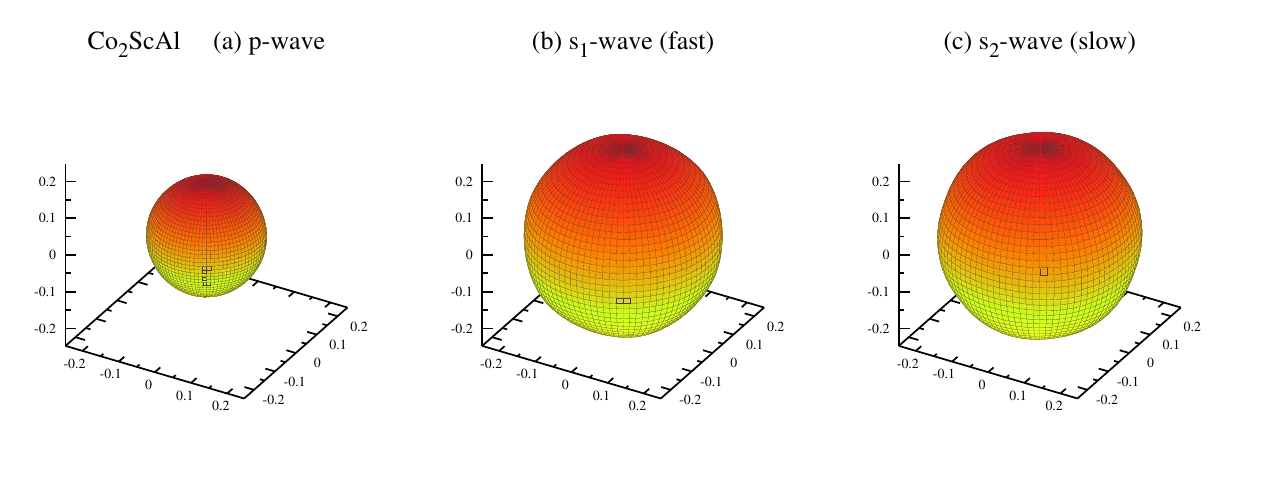}
\includegraphics[width=15cm]{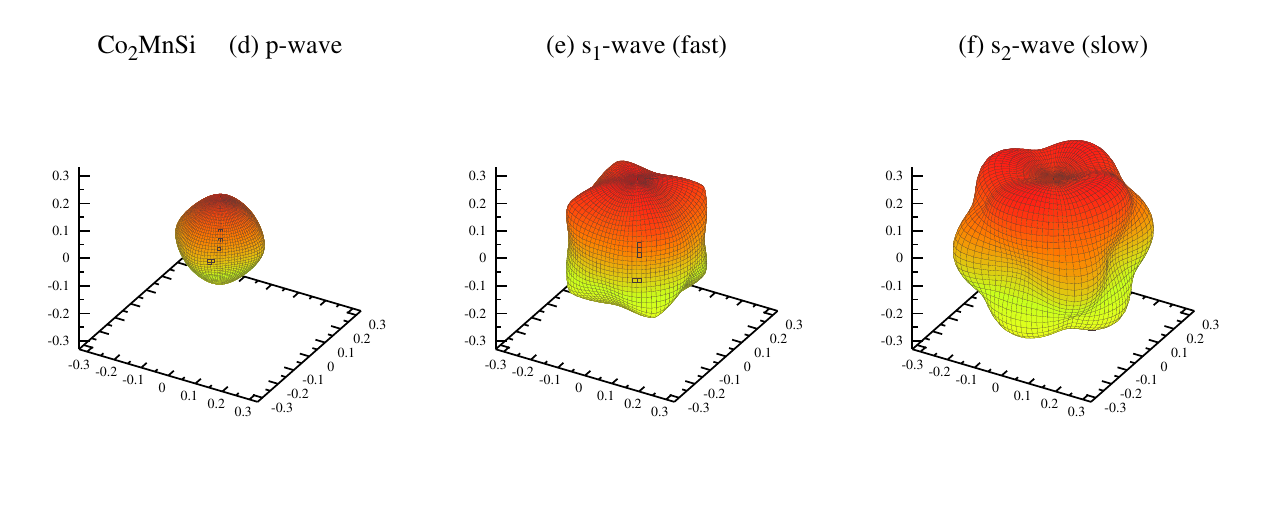}
   \caption{(Online color) Calculated slowness surfaces of 
            Co$_2$ScAl (a)...(c) and Co$_2$MnSi (d)...(f). \\
            The slowness is given in (km/s)$^{-1}$. }
\label{fig:slow_compare}
\end{figure*}

Figure~\ref{fig:slow_compare} compares the slowness surfaces of Co$_2$ScAl and
Co$_2$MnSi. The latter was considered, since it is known from  experiments to be very
stable and to exhibit very low disorder. Further, it is the Heusler material with
highest tunneling magneto resistance (TMR) ratios. The isotropic elastic
behavior of Co$_2$ScAl is reflected in the nearly spherical distributions
describing its three slowness surfaces. The shear modes are nearly degenerate
and exhibit {\it kiss singularities} in the high symmetry directions.
The behavior of the slowness surfaces of Co$_2$MnSi is typical for most of the
investigated Co$_2$-based Heusler compounds, and its shape is similar to that of
Cu$_2$MnAl. Its lower asymmetry as compared to Cu$_2$MnAl results in less
pronounced differences between minima and maxima of the slowness.

% Figure 12 %%%%%%%%%%%%%%%%%%%%%%%%%%%%%%%%%%%%%%%%%%%%%%%%%%%%%%%%%%%%%%%%%%%%%
%%%%%%%%%%%%%%%%%%%%%%%%%%%%%%%%%%%%%%%%%%%%%%%%%%%%%%%%%%%%%%%%%%%%%%%%%%%%%%%%

To use slowness as a parameter for determination of the elastic constants,
measurements along different high symmetry directions may be used. For the
pressure wave, it may be found, for example, the following slowness eigenvalues
$s$: $s_p^{001}=\sqrt{\rho/c_{11}}$, $s_p^{110}=\sqrt{\rho/c_{44}}$, and
$s_p^{111}=\sqrt{3\rho/(c_{11}+4c_{44}+2c_{12})}$ (superscript indices indicates
the high symmetry direction [hkl], subscript index indicates the direction of
polarization). The shear waves are degenerate in the [001] and [111] directions
with values $s_{s1,s2}^{001}=\sqrt{\rho/c_{44}}$, and
$s_{s1,s2}^{111}=\sqrt{3\rho/(c_{11}+c_{44}-c_{12})}$, respectively. The two non
degenerate values for the [110] direction are $s_{s1}^{110}=\sqrt{2\rho/(c_{11}-
c_{12})}$ and $s_{s2}^{110}=\sqrt{2\rho/(c_{11}+2c_{44}+c_{12})}$.

%%%%%%%%%%%%%%%%%%%%%%%%%%%%%%%%%%%%%%%%%%%%%%%%%%%%%%%%%%%%%%%%%%%%%%%%%%%%%%%%
\section{Summary}

The elastic and accompanying physical properties of Heusler compounds, Cu$_2$MnAl
as well as the Co$_2YZ$ family, have been calculated. The present results for
Cu$_2$MnAl are in good agreement with experiments. The directions of the largest
Young's modulus indicates the high-fracture energy directions. According to the
Bader's QTAIM analysis, the bonding in Cu$_2$MnAl is clearly metallic. However,
Cu$_2$MnAl has one of the highest anisotropies as compared to the Co$_2$-based
compounds. The Debye temperature, where only acoustic vibrational modes
contribute, is about 397~K, which is lower than that of most of the studied
compounds based on Co$_2$.

Based on the calculation of their elastic properties, the crystalline stability
of Co$_2$-based cubic Heusler materials was assessed. The elastic constants of
all the studied compounds follow the general inequality $B > c_{44} > G > c' > 0$
such that the rigidity modulus $G$ is the main constraint on stability. The
results of  our calculations demonstrate that all the studied compounds are
close to the borderline between brittle and ductile. From the elastic point of
view, they mainly exhibit bonding behavior between those of covalent and
metallic. For most of the studied stable compounds, the universal anisotropy
index is in the range $0.57<A_U<2.73$. All the studied stable compounds are most
stiff in the $\left<111\right>$-type directions. The detailed analysis of all
the compounds revealed that Pugh's criterion for the ductile--brittle transition
($k^{\rm Pugh}_{\rm cr}\approx1.75$) should be replaced by Christensen's
criterion ($k_{\rm cr}\approx2.27$ or $k_{\rm cr}^{-1}\approx0.44$).

For Co$_2$-based compounds, when the nuclear charge $Z$ of the main group
element increases, the lattice parameters also increase but the values of the
hardness parameter decrease. Here, the Zener  ratio ($A_e$) and the universal
anisotropy index ($A_U$) show the same tendency. It should be mentioned however,
that the anisotropy does only judge on the structural stability but not on the
chemical stability, as is seen especially for the case of Co$_2$MnSb, that does
not exist as a pure compound. The Debye temperature in the acoustical approach
decreases with $Z$, but there is no evidence for a distinguishable dependence on
the $T$ element for the Co$_2TM$ compounds. The Gr{\"u}neisen parameters in the
acoustical approach are all nearly the same for the different compounds. The
hardness shows the same tendency. Finally, it is found that Co$_2$ScAl is close
to an ideal Cauchy solid and is predicted to be the most hard material in the
investigated series.

The calculated material properties can be applied quite reliably to bulk
materials. On the other hand, the prediction of stability could be exploited in
any compounds in particular to those when the possibility of structural phase
transition in crystals is investigated.

%%%%%%%%%%%%%%%%%%%%%%%%%%%%%%%%%%%%%%%%%%%%%%%%%%%%%%%%%%%%%%%%%%%%%%%%%%%%%%%%
%\begin{acknowledgments}

%The authors gratefully acknowledge themselves for providing coffee and chocolate ... .

%\end{acknowledgments}
%%%%%%%%%%%%%%%%%%%%%%%%%%%%%%%%%%%%%%%%%%%%%%%%%%%%%%%%%%%%%%%%%%%%%%%%%%%%%%%%

%%%%%%%%%%%%%%%%%%%%%%%%%%%%%%%%%%%%%%%%%%%%%%%%%%%%%%%%%%%%%%%%%%%%%%%%%%%%%%%%
\appendix 
\label{appendix}
\section{Cubic elastic constants, elastic moduli, and elastic stability.}

The basics of the elastic properties of solids are described in the book by 
Nye~\cite{Nye85}. Here the focus is on the equations for cubic compounds,
remarks on tetragonal and hexagonal compounds are found in Reference~\cite{WNF17}.
A lattice $A$ can be transformed to a new deformed lattice 
$A'$ by the strain matrix $\epsilon$. The strain matrix $\epsilon$ is symmetric 
and contains six different strains $e_i$. By Hooke's law, the elastic relation 
between strain ($\epsilon$) and stress ($\sigma$) is:

\begin{equation}
  \sigma = {\bm C} \epsilon,
\end{equation}

where ${\bm C}$ is the elastic stiffness matrices, and the relations between the 
compliance matrix ($\bm S$) and the stiffness matrix is

\begin{equation}
  {\bm S} = {\bm C}^{-1}.
\end{equation}

In the most general case, the elastic matrix is symmetric and on the order of
$6 \times6$~\cite{Nye85}. In triclinic lattices, it contains 21 independent
elastic constants. This number is largely reduced in highly symmetric lattices.
In cubic lattices, only the three elastic constants $c_{11}$, $c_{12}$, and
$c_{44}$ are independent. The elastic matrix for all classes of cubic crystals
has the form (zero elements are denoted by dots):

\begin{equation}
  {\bm C}^{\rm cubic} = \left(
  \begin{array}{ccc ccc} 
        c_{11}   & c_{12}    & c_{12} &  .      & .      & . \\
        c_{12}   & c_{11}    & c_{12} &  .      & .      & . \\
        c_{12}   & c_{12}    & c_{11} &  .      & .      & . \\
        .        & .         & .      &  c_{44} & .      & . \\
        .        & .         & .      &  .      & c_{44} & . \\
        .        & .         & .      &  .      & .      & c_{44}  \\                                      
  \end{array}
  \right)
\end{equation}

The matrix has 6 eigenvalues out of which only three are different.
These three different values o the eigenvalues are:
\begin{itemize}
	\item  $C_1^c = 2c_{12}+c_{11}$ (nondegenerate), 
	\item  $C_{2,3}^c = c_{11}-c_{12}$ (twofold), and 
	\item  $C_{4,\ldots,6}^c = c_{44}$ (threefold degenerate).
\end{itemize}
They correspond to the bulk, the tetragonal shear, and the shear 
moduli as will be shown below. The crystal becomes unstable
when one of the eigenvalues becomes zero or negative.

For an isotropic system, the elastic matrix $C^{\rm iso}$ contains only the two 
constants $c_{11}$ and $c_{12}$, whereas the remaining diagonal elements of the 
matrix are determined by $c_{44}=(c_{11}-c_{12})/2$.

In cubic systems, the relations between the elastic constants $c_{ij}$ and the 
elements of the compliance matrix $s_{ij}$ are given by,

\begin{eqnarray}
  s_{11} & = & \frac{c_{11} + c_{12}}{c}            \\ \nonumber 
  s_{12} & = & \frac{- c_{12}}{c}                   \\ \nonumber 
       c & = & (c_{11} - c_{12})(c_{11} + 2 c_{12}) \\ \nonumber 
  s_{44} & = & \frac{1}{c_{44}}
\end{eqnarray}

The bulk modulus is defined by the elastic constants.
For cubic materials it is given by,

\begin{equation}
  B = \frac{1}{3}(c_{11} + 2c_{12}).
\label{eq:BM}
\end{equation}

Born and co-workers developed the theory of stability of crystal
lattices~\cite{Bor40,*Mis40,*BFu40,*BMi40,*Fue41a,*Fue41b}. 
The Born--Huang~\cite{BHu56} elastic stability criteria for a cubic crystal at
ambient conditions~\cite{WVP93} are given by,

\begin{itemize}
  \item $c_{11} + 2c_{12} > 0$
  \item $c_{44} > 0$
  \item $c_{11} - c_{12} > 0$
\end{itemize}

that is, the bulk, $c_{44}$-shear, and tetragonal shear moduli have to be all 
positive. The criteria are referred to as spinodal, Born's shear~\cite{Bor39} 
and Born criteria, respectively. The first criterion defines the spinodal 
pressure,

\begin{equation}
   p_s = c_{11} + 2c_{12},
\end{equation}

whereas the last criterion is often used to define an additional elastic constant,

\begin{equation}
   C' = c_{11} - c_{12}.
\end{equation}

This constant is also called tetragonal shear modulus. In some studies, 
$c' = (c_{11}-c_{12})/2$
is used instead, because the tetragonal instability is 
observed when the hydrostatic pressure becomes $2p>C'$, that is $p>c'$. In 
detail, $G_{110} = c'$ is the single-crystal shear modulus for the (110) plane 
along the [110] direction. The single-crystal shear modulus for the (100) plane 
along the [010] direction is $G_{100} = c_{44}$. It is related to a tetragonal 
deformation and large values denote high stability of the crystal with respect 
to tetragonal shear.

The Cauchy pressure for cubic crystals is defined using the Cauchy relation as,

\begin{equation}
   p_{\rm C} = c_{12} - c_{44}.
\end{equation}

For single cubic crystals, the shear modulus $G$, Pugh's ratio $k=B/G$, and the
Poisson's ratio $\nu$ are calculated from the elastic constants using the
following relations:

\begin{eqnarray}
   G    & = & \frac{3 c_{44} + c_{11} - c_{12} }{5}                                \\ \nonumber
   k    & = & \frac{5}{3} \frac{(c_{11} + 2c_{12})}{(3 c_{44} + c_{11} - c_{12})}  \\ \nonumber
   \nu  & = & \frac{c_{12} }{c_{11} + c_{12}}.
\end{eqnarray}

The first and third Born criteria restrict the range of Poisson's ratio to 
$-1\leq\nu\leq1/2$.

Polycrystalline materials consist of randomly oriented crystals and thus a
description of their elastic properties requires only two independent elastic
moduli: the bulk modulus ($B$), and the shear modulus ($G$). The relationships
between the single-crystal elastic constants and the polycrystalline elastic
moduli are given by the Voigt~\cite{Voi28} or Reu{\ss}~\cite{Reu29} averages.
Voigt's approach uses the elastic stiffnesses $c_{ij}$, whereas Reu{\ss}'s approach
uses the compliances $s_{ij}$. The bulk moduli in Voigt's ($B_V$) and Reu{\ss}'s
($B_R$) approach are equal for cubic crystals and given by:

\begin{eqnarray}
B & = & \frac{1}{3}(c_{11} + 2c_{12})            \\ \nonumber
  & = & \frac{1}{3}\frac{1}{(s_{11} + 2s_{12})}  \\ \nonumber
  & = & B_V = B_R.
\label{eq:eos}
\end{eqnarray}

The isotropic shear or rigidity modulus $G = [G_V + G_R]/2$ 
is defined by Voigt's $G_V$ and Reu{\ss}'s $G_R$ shear moduli, where,

\begin{eqnarray}
	G_V & =  & \frac{1}{5} ( c_{11} - c_{12} + 3c_{44} )   \\ \nonumber
	G_R & =  & \frac{5}{4(s_{11} - s_{12}) + 3 s_{44}}                \\ \nonumber
	    & =  & 5 \frac{(c_{11} - c_{12})c_{44}}{3(c_{11} - c_{12}) + 4 c_{44}}.
\end{eqnarray}

Accordingly, Poissons's ratio $\nu$ and Young's modulus $E$ of polycrystalline 
cubic materials are calculated from the equations using the averaged bulk and 
rigidity moduli as,

\begin{eqnarray}
     \nu & = & \frac{1}{2} \ \frac{3B - 2G}{3B + G}    \\ \nonumber
     E   & = & 2 G (1+\nu).
\end{eqnarray}

In cubic crystals, the bulk modulus is isotropic. However,  rigidity and Young's
moduli not isotropic. The directional dependence of Young's modulus $E(\hat{r})$
is defined by the ratio of longitudinal stress to strain. For cubic systems, the
three dimensional distribution is given by,

\begin{equation}
  E^{-1}(\hat{r}) = s_{11} - 2s F_{lmn},
\end{equation}
 
where $s=s_{11} - s_{12} - s_{44}/2$, and
$F_{lmn}=(\hat{x}^2\hat{y}^2 + \hat{y}^2\hat{z}^2 + \hat{z}^2\hat{x}^2)$ 
is the orientation function of a cubic single crystal specimen given in terms 
of the direction cosines ($l:=\hat{x}$, etc.). It is obvious that $E(\hat{r})$ 
becomes isotropic for $s=0$. Hence, Zener ratio or the elastic 
anisotropy is defined for cubic crystals as, 

\begin{equation}
  A_e = \frac{2 (s_{11} - s_{12})}{s_{44}} = \frac{2 c_{44}}{c_{11} - c_{12}}.
\end{equation}

The cubic elastic anisotropy may be used as another important physical quantity 
for the description of  structural stability. Materials exhibiting 
large $A_e$ ratios occasionally show a tendency to deviate from the cubic 
structure. Materials with negative Zener ratio ($A_e<0$) violate at least one 
of Born's criteria and are mechanically instable.

Ranganathan and Ostoja-Starzewski~\cite{ROs08} summarized the existing
anisotropy theories and developed a so-called universal anisotropy index $A_U$
that is calculated for cubic crystals, using the condition $B_V = B_R$ by the
simplified equation,

\begin{eqnarray}
A_U          & = & 5 \frac{G_V}{G_R}+\frac{B_V}{B_R}-6     \\ \nonumber
A_U^{cubic}  & = & 5 \left( \frac{G_V}{G_R} .– 1 \right)
\end{eqnarray}

Similar to the case of the Young's modulus, the directional dependence of the
rigidity modulus $G(\hat{r})$ is defined by~\cite{Goo41},

\begin{eqnarray}
\label{eq:GR}
  G_0^{-1}(\hat{r}) & = & s_{44} + 4s F_{lmn}, \\ 
  G^{-1}(\hat{r})   & = & G_0^{-1}(\hat{r}) - \frac{2s^2(F_{lmn}-4F_{lmn}^2+3\chi_{lmn})}{s_{11}-4sF_{lmn}}, \nonumber
\end{eqnarray}

where $\chi_{lmn}=\hat{x}^2\hat{y}^2\hat{z}^2$. The last term in
Eq. ~(\ref{eq:GR}) is the so-called bending--torsion correction (or
difference) if $G_0$ is defined as the {\it "true"} rigidity
modulus~\cite{Goo41}. $G_0$ becomes isotropic for $A_e=1$. The bending--torsion
correction vanishes for the highly symmetric $\left<100\right>$,
$\left<110\right>$, and $\left<111\right>$-type directions.

The Cauchy criterion of vanishing Cauchy pressure for crystals with cubic 
symmetry is $c_{12}=c_{44}$. The conditions required to satisfy this Cauchy 
relation are:
\begin{itemize}
	\item Only central forces take part in the interaction between the atoms.
	\item Only harmonic forces exist between the atoms. Anharmonicity will
        destroy the Cauchy relations. 
	\item The atoms are located at the centers of symmetry.
	\item Thermal effects and initial stress are neglected.
\end{itemize}

From the isotropy ($c_{12}=c_{11}-2c_{44}$) and Cauchy ($c_{12}=c_{44}$) 
relations, only one independent elastic constant ($c_{11}=3c_{12}=3c_{44}$) 
would remain for cubic crystals. This has the result that Pugh's ratio of a 
cubic, isotropic solid following Cauchy's relation becomes 
$k_{\rm Cauchy}=5/3=1.6\overline{6}\approx1.7$. At the same time, Poison's ratio
simplifies to $\nu_{\rm Cauchy}=1/4$. The elastic matrix of such an 
ideal {\it Cauchy solid} has the form (zero elements are denoted by dots),

\begin{equation}
  {\bm C}^{\rm cubic}_{\rm Cauchy} = \frac{1}{3}\left(
  \begin{array}{lll ccc} 
        3 c_{11} & c_{11}   & c_{11}   &  .       & .      & . \\
        c_{11}   & 3 c_{11} & c_{11}   &  .       & .      & . \\
        c_{11}   & c_{11}   & 3 c_{11} &  .       & .      & . \\
        .        & .        & .        & c_{11}   & .      & . \\
        .        & .        & .        &  .       & c_{11} & . \\
        .        & .        & .        &  .       & .      & c_{11}. \\                                      
  \end{array}
  \right)
\end{equation}

The three different eigenvalues of ${\bm C}^{\rm cubic}_{\rm Cauchy}$ are
$5c_{11}$, $2c_{11}$, and $c_{11}$, which are nondegenerate, twofold degenerate,
and threefold degenerate, respectively.

Apart from the elastic moduli, a few more important physical quantities can be derived
from the elastic constants. The volume ($\kappa$) and linear ($\beta$)
compressibilities of cubic crystals are isotropic and given by,

\begin{eqnarray}
  \beta  & = & s_{11} + 2s_{12}, \\ \nonumber
  \kappa & = & 3 \beta = B^{-1}.
\end{eqnarray}

% %%%%%%%%%%%%%%%%%%%%%%%%%%%%%%%%%%%%%%%%%%%%%%%%%%%%%%%%%%%%%%%%%%%%%%%%%%%%%

\section{Derived physical properties from cubic elastic constants.}

In the bond-orbital model,  Kleinman's internal displacement parameter is 
defined by~\cite{Harr89}:

\begin{equation}
  \zeta = \frac{c_{11} + 8 c_{12}}{7 c_{11} + 2 c_{12}}.
\end{equation}

It describes the relative positions of atoms in different sublattices under 
volume conserving strain distortions for which the positions are not fixed by 
symmetry anymore. $\zeta$ vanishes if no internal displacements appear. 
$\zeta=1$ when the bond lengths are unchanged and $\zeta=-1/2$ when the bond 
angles are unchanged, both for linear strain. 

In the quasi-harmonic approach, the Debye temperature $\Theta^{qha}_{D}$ depends 
upon the volume of the crystal. For every volume $V$, $\Theta^{qha}_{D}(V)$ is 
rigorously defined in terms of the elastic constants through a spherical average 
of the three components of the sound velocity. The isotropic approximation, 
allows to evaluate $\Theta^{qha}_{D}$ using the expression~\cite{REd66,Poi91}, 

\begin{eqnarray}
  \Theta^{qha}_{D}           & = \frac{\hbar}{k_B}\sqrt[3]{n \cdot 6 \pi^2 \sqrt{V}} \sqrt{\frac{B}{M}}f_\nu  , \\ \nonumber
  \Theta_{qha}^{\rm Heusler} & = \frac{\hbar}{k_B}\sqrt[3]{96 \pi^2} \sqrt{a \frac{B}{M}}f_\nu ,
\end{eqnarray}

where $\hbar$ is Planck's constant, $k_B$ is Boltzmann's constant and $n$ is the 
number of atoms in the primitive cell with volume $V$ unit ($n=4$ in the case of 
Heusler compounds and the volume of the primitive cell or 16 for the cubic cell 
with lattice parameter $a$), $B$ is the adiabatic bulk modulus of the crystal 
and $M$ the mass of the compound corresponding to $V$. Finally, $f_\nu$ is a 
function of the Poisson ratio $\nu$~\cite{Fra98}:

\begin{equation}
  f_\nu =  \sqrt[3]{ \frac{3}{ 2 \left[ \frac{2}{3}  \frac{1+\nu}{1-2\nu} \right]^{3/2}   +   \left[ \frac{1}{3}  \frac{1+\nu}{1-\nu} \right]^{3/2} } }.
\end{equation}

% %%%%%%%%%%%%%%%%%%%%%%%%%%%%%%%%%%%%%%%%%%%%%%%%%%%%%%%%%%%%%%%%%%%%%%%%%%%%%

Another important mechanical property is the hardness of a 
material~\cite{Gil09}. Other than for the elastic moduli, there is no 
straightforward theory to calculate the hardness directly from the elastic 
constants. However, several models were developed to relate the hardness of a 
material to the elastic moduli. Pugh~\cite{Pug54} related the Brinell hardness 
$H_B$ of pure metals to their shear modulus $G$ by $H_B=G\:b/c$, where $b$ is 
the Burger's vector of the dislocation and $c$ is a constant for all metals of 
the same structure. Teter~\cite{Tet98} obtained the semi-empirical relation 
$H_V^{\rm T}\approx0.151 G$ between Vicker's hardness $H_V$ and the rigidity 
modulus $G$. Recently, Chen {\it et al}~\cite{CNL11} gave a semi-empirical 
relation between Vicker's hardness and the product of the squared Pugh's modulus 
($k=B/G$) ratio and the shear modulus $G$ as,

\begin{equation}
  H{_V\rm ^C} = 2(k^{-2} G)^{0.585} - 3.
\end{equation}

% %%%%%%%%%%%%%%%%%%%%%%%%%%%%%%%%%%%%%%%%%%%%%%%%%%%%%%%%%%%%%%%%%%%%%%%%%%%%%

Also for cubic metals only, Fine et al.~\cite{Fine84} obtained an approximate linear 
relationship between the melting temperature $T_m$ and the elastic constant 
$c_{11}$. The $T_m^c$ of various cubic metals was  $\pm300$~K within a linear 
dependence when estimated from the following empirical equation:

\begin{equation}
	T_m^c = 553 \: {\rm K} + 5.91 \frac{\rm K}{\rm GPa} \cdot c_{11}.
\end{equation}

% %%%%%%%%%%%%%%%%%%%%%%%%%%%%%%%%%%%%%%%%%%%%%%%%%%%%%%%%%%%%%%%%%%%%%%%%%%%%%

The elastic constants and moduli also allow estimation of the averaged sound 
velocity $\overline{v}$,

\begin{equation}
  \overline{v} = \left[ \frac{3}{v_l^{-3} + 2 v_t^{-3}} \right] ^{1/3}.
\end{equation}

From the longitudinal ($v_l$) and transverse ($v_t$) 
elastic wave velocities of isotropic materials the Debye 
temperature  can be estimated, where $v_l$ and $v_t$ are, 

\begin{eqnarray}
  v_l & =  & \sqrt{ \frac{3B+4G}{3\rho} }   \\ \nonumber
  v_t & =  & \sqrt{ \frac{G}{\rho} },
\end{eqnarray}

where, $\rho$ is the mass density of the material. From the average sound velocity at
low temperatures, the Debye temperature can be estimated by using the
relation~\cite{And63}:

\begin{equation}
  \Theta^{\rm ac}_D = \overline{v} \frac{h}{k_B} \sqrt[3]{ \frac{n}{4\pi} \frac{N_A \rho}{M} }
                    = \overline{v} \frac{h}{k_B} \sqrt[3]{ \frac{n}{4\pi} \frac{1}{V} } ,
\end{equation}

where $N_A$ is Avogadoro's number. Other parameters are the same
as in the case of the Debye temperature calculation in the quasi-harmonic approach.

In solids, the Gr{\"u}neisen parameter $\zeta^{ac}$ is also related to the sound
velocities. Belomestnykh~\cite{Bel04} derived this Gr{\"u}neisen parameter using,

\begin{equation}
  \zeta^{\rm ac} = \frac{3}{2} \frac{(3v_l^2 - 4 v_t^2)}{(v_l^2 + 2v_t^2)}.
\end{equation}

% %%%%%%%%%%%%%%%%%%%%%%%%%%%%%%%%%%%%%%%%%%%%%%%%%%%%%%%%%%%%%%%%%%%%%%%%%%%%%

The above described acoustical properties concern averages and may be used for
polycrystalline materials. Acoustical spectroscopy is used, indeed, also for
investigation of the single crystal elastic constants. The directional
dependence of the phase velocity $v$ is found from Christoffel's equation:

\begin{equation}
	(\Gamma_{ij} - \rho v^2 \delta_{ij}) U_j = 0,
\end{equation}

where $\Gamma_{ij}=c_{ijkl}l_jl_l$ is the Christoffel tensor built from the
elastic constants and the direction cosines $l_i$ ($i=1\ldots3$). $\rho$ is the
density, $\delta_{ij}$ is the Kronecker symbol and $U$ is the polarization
vector.
 
In cubic systems, only 3 elastic constants are independent and the components of
the Christoffel tensor are reduced to,

\begin{equation}
	\Gamma_{ij} = \left\{
                 \begin{array}{ll}
	               (c_{11}+c_{44})l_il_j         & i \neq j \\
	               c_{11}l_i^2 + c_{44}(1-l_i^2) & i = j    \\
                 \end{array}
                \right.
\end{equation}

with $l_1=\sin(\theta)\cos(\phi)$, $l_2=\sin(\theta)\sin(\phi)$, and
$l_3=\cos(\theta)$ in polar co-ordinates. In greater detail, the elements of the
Christoffel tensor are (note that: $\Gamma_{ij}=\Gamma_{ji}$),

\begin{eqnarray}
 \Gamma_{11} & = & c_{11} l_1^2 + c_{44} l_2^2 + c_{44} l_3^2   \\  \nonumber
 \Gamma_{22} & = & c_{44} l_1^2 + c_{11} l_2^2 + c_{44} l_3^2   \\  \nonumber
 \Gamma_{33} & = & c_{44} l_1^2 + c_{44} l_2^2 + c_{11} l_3^2   \\  \nonumber
 \Gamma_{12} & = & (c_{12} + c_{44}) l_1l_2   \\  \nonumber
 \Gamma_{13} & = & (c_{12} + c_{44}) l_1l_3   \\  \nonumber
 \Gamma_{23} & = & (c_{12} + c_{44}) l_2l_3
\end{eqnarray}

In case of an ideal Cauchy solid the
Christoffel tensor is further reduced to,

\begin{equation}
  {\bm \Gamma} = \frac{c_{11}}{3}  \left(
  \begin{array}{ccc} 
        2 l_1^2  +  1   & 2 l_1l_2    & 2 l_1l_3    \\
        2 l_1l_2        & 2 l_2^2 + 1 & 2 l_2l_3    \\
        2 l_1l_3        & 2 l_2l_3    & 2 l_3^2 + 1. \\                                     
  \end{array}
  \right)
\end{equation}

The results for ideal Cauchy solids are three eigenvalues: $\rho v_p^2=c_{11}$
for the compression wave, and $\rho v_{s1, s2}^2=c_{11}/3$ for the twofold
degenerate shear wave. Both the modes, shear $s$ and pressure $p$, are independent
of the propagation direction and their slowness surfaces appear spherical with
$s_p=\sqrt{\rho/c_{11}}$ and $s_s=\sqrt{3\rho/c_{11}}$.

% %%%%%%%%%%%%%%%%%%%%%%%%%%%%%%%%%%%%%%%%%%%%%%%%%%%%%%%%%%%%%%%%%%%%%%%%%%%%%

\bibliography{elastic_cubic}

\end{document}